\documentclass[traditabstract]{aa}
\usepackage{epsfig}
\usepackage{txfonts}
\usepackage{natbib}
\bibpunct{(}{)}{;}{a}{}{,}

\newcommand{\target}{KIC\,11558725}
\newcommand{\kep}{{\em Kepler}}

\newcommand{\porb}{$P_{\rm{orb}}$}
\newcommand{\teff}{\ensuremath{T_{\rm{eff}}}}
\newcommand{\logg}{\ensuremath{\log g}}

\newcommand{\lheh}{\ensuremath{\log \left(N_{\mathrm{He}}/N_{\mathrm{H}}\right)}}
\newcommand{\ua}{\ensuremath{\uparrow}}
\newcommand{\da}{\ensuremath{\downarrow}}

\begin{document}



\title{Three ways to solve the orbit of \target: a 10 day beaming
  sdB+WD binary with a pulsating subdwarf\thanks{{ Based on observations
    obtained by the \kep\ spacecraft, the Kitt Peak Mayall Telescope, 
    the Nordic Optical Telescope and the William Herschel Telescope}}}

\author{
       J.~H.~Telting   \inst{1}  
 \and  R.~H.~\O stensen\inst{2}
 \and  A.~S.~Baran     \inst{3}
 \and  S.~Bloemen      \inst{2}
 \and  M.~D.~Reed      \inst{3}
 \and  R.~Oreiro       \inst{4}
 \and  L.~Farris       \inst{3}
 \and  T.~A.~Ottosen   \inst{5}
 \and  C.~Aerts        \inst{2,6}
 \and  S.~D.~Kawaler   \inst{7}
 \and  U.~Heber        \inst{8}
 \and  S.~Prins        \inst{2}
 \and  E.~M.~Green     \inst{9}
 \and  B.~Kalomeni     \inst{10}
 \and  S.~J.~O'Toole   \inst{11}
 \and  { F.~Mullally}     \inst{12}
 \and  { D.~T.~Sanderfer} \inst{12}
 \and  { J.~C.~Smith}     \inst{12}
 \and  { H.~Kjeldsen}     \inst{5}
}

\institute{
Nordic Optical Telescope, Apartado 474, E-38700 Santa Cruz de La Palma, 
Spain\\     e-mail: {\tt jht@not.iac.es}
\and
Instituut voor Sterrenkunde, KU Leuven, Celestijnenlaan 200D,
B-3001 Leuven, Belgium
\and
Department of Physics, Astronomy, and Materials Science,
Missouri State University, Springfield, MO 65804, USA
\and
Instituto de Astrof\'{i}sica de Andaluc\'{i}a (CSIC), Glorieta de la
Astronom\'{i}a s/n, E-18008 Granada, Spain
\and
Department of Physics and Astronomy, Aarhus University, DK-8000 Aarhus C, Denmark
\and
Department of Astrophysics, IMAPP, Radboud University Nijmegen, PO Box
9010, NL-6500 GL Nijmegen, the Netherlands
\and
Department of Physics and Astronomy, Iowa State University, Ames, IA 50011 USA
\and
Dr. Remeis Sternwarte Bamberg, Universit\"at  Erlangen-N\"urnberg, Germany
\and
Steward Observatory, University of Arizona, 933 North Cherry Avenue, 
Tucson, AZ 85721, USA
\and 
University of Ege, Department of Astronomy \& Space Sciences, 
35100 {\.I}zmir, Turkey
\and 
Australian Astronomical Observatory, PO Box 296, Epping, NSW 1710, Australia
\and 
{SETI Institute/NASA Ames Research Center, Moffett Field, CA 94035}
}

\date{submitted 23/04/2012 ; accepted 16/06/2012}

\titlerunning{KIC\,11558725: a 10 day beaming sdBV+WD binary}
\authorrunning{J.~H.~Telting et al.}

\abstract{ 
  The recently discovered subdwarf B (sdB) pulsator \target\ is one of the 16
  pulsating sdB stars detected in the \kep\ field.  It features a
  rich $g$-mode frequency spectrum, with a few low-amplitude $p$-modes
  at short periods. This makes it a promising target for a seismic
  study aiming to constrain the internal structure of this star, and
  of sdB stars in { general.  

  We have obtained ground-based spectroscopic radial-velocity measurements of
  \target\ based on low-resolution spectra in the Balmer-line region,
  spanning the
  2010 and 2011 observing seasons.  From these data we have discovered
  that \target\ is a binary with period $P$=10.05\,d, and that the
  radial-velocity amplitude of the sdB star is 
  58\,km\,s$^{-1}$. Consequently the companion of the sdB star has a
  minimum mass of 0.63\,$M_{\sun}$, and is therefore most likely 
  an unseen white dwarf.  

  We analyse the near-continuous 2010--2011 \kep\ light curve to
  reveal the orbital Doppler-beaming effect, giving rise to light
  variations at the 238\,ppm level, which is consistent with the
  observed spectroscopic orbital radial-velocity amplitude of the
  subdwarf.

  We use the strongest 70 pulsation frequencies in the \kep\ light
  curve of the subdwarf as clocks to derive a third consistent
  measurement of the orbital radial-velocity amplitude, from the
  orbital light-travel delay.  The orbital radius $a_{\rm sdB}\sin
  i$\,=\,11.5\,$R_{\sun}$ gives rise to a light-travel time delay of
  53.6\,s, which causes aliasing and lowers the amplitudes of the
  shortest pulsation frequencies, unless the effect is corrected for.

  We use our high signal-to-noise average spectra to study the atmospheric
  parameters of the sdB star, deriving \teff\,=\,27\,910\,K and
  \logg\,=\,5.41\,dex, and find that carbon, nitrogen and oxygen are
  underabundant relative to the solar mixture.
 
  Furthermore, we analyse the \kep\ light curve for its pulsational
  content and extract more than 160 significant frequencies.  }
  We investigate the pulsation frequencies for expected period
  spacings and rotational splittings.  We find period-spacing
  sequences of spherical-harmonic degrees $\ell$=1 and $\ell$=2, and
  we associate a large fraction of the $g$-modes in \target\ with
  these sequences.

  From frequency splittings we conclude that
  the subdwarf is rotating subsynchronously with respect to the orbit.
}


\maketitle

\section{Introduction}

The hot subdwarf\,B (sdB) stars populate an extension of the horizontal
branch where the hydrogen envelope is of too low a mass to sustain hydrogen
burning. These core helium burning stars must have suffered extensive
mass loss close to the tip of the red giant branch in order to reach
this core/envelope configuration. Binary interactions, either through
stable Roche lobe overflow or common envelope ejection, are likely
to be responsible for the majority of the sdB population
\citep[see][for a detailed review]{heber09}.

\begin{figure}[t]
\centerline{\psfig{figure=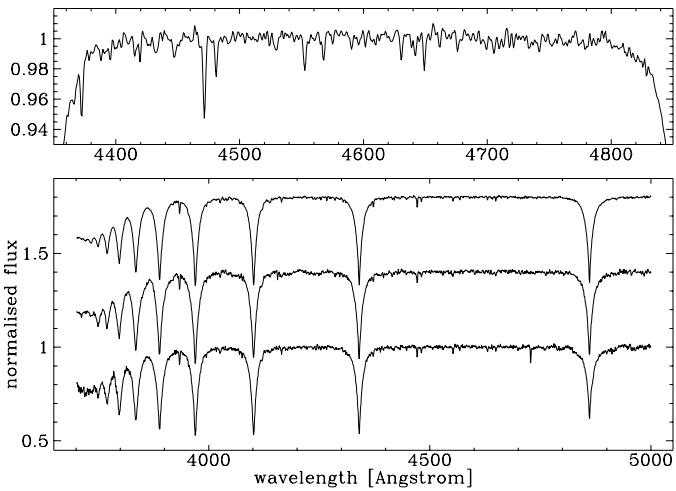,width=6.5cm,angle=-90}}
\caption[]{Mean spectra from KP4m, NOT, WHT (bottom to top), offset in
  flux for clarity.   The top panel is a zoom-in of the WHT
  spectrum, demonstrating some of the stronger lines of heavy
  elements, such as MgII 4481\,\AA\ and SiIII 4552,\,4567\,\AA. }
\label{fig:meanSpectra}
\end{figure}

Several extensive radial-velocity surveys have targeted the sdB stars,
with the most recent large sample explored by \citet{copperwheat11}.
They find that $\sim$50\%\ of all sdB stars reside in short-period
binary systems with the majority of companions being white dwarf (WD)
stars. A recent compilation of such short-period systems can be found
in Appendix A of \citet{geier11a}, which lists a total of 89
systems. Adding 18 new systems from \citet{copperwheat11} brings the
total well above a hundred sdBs with periods ranging from 0.07 to
27.8\,d. These systems are all characterised by being single-lined
binaries, {\em i.e.} only the sdB stars contribute to the optical
flux, which directly constrains the companion to be either an M-dwarf
or a compact stellar-mass object.  Binaries with companions of type
earlier than M are double-lined and also readily identifiable from a
combination of optical and infrared photometry. \citet{reed04} find
that $\sim$50\%\ of all sdB stars have IR excess and must have a
companion no later than M2. Radial velocity studies targeting these
double-lined stars have had a hard time detecting orbital periods,
indicating that they must be exceedingly long. A recent breakthrough
was made by \citet{ostensen12a} using high-resolution spectroscopy of
a sample of eight bright subdwarf + main-sequence (MS) binaries
detecting orbital periods spanning a range from $\sim$500 to 1200\,d
with velocity amplitudes between 2 and 8\,km\,s$^{-1}$. The period
distribution of these different types of binary systems are important
in that they can be used to constrain a number of vaguely defined
parameters used in binary population synthesis models, including the
common envelope ejection efficiency, the mass and angular momentum
loss during stable mass transfer, the minimum core mass for helium
ignition, etc.  The seminal binary population study of
\citet{han02,han03} successfully predicts many aspects of the sdB
star population, but the key parameters have a wide range of possible
values. A recent population synthesis study by \citet{clausen12}
explores the possible populations of sdB+MS stars and demonstrates how
the entire population can change with different parameter sets, but
does not deal with sdB+WD binaries.

A theoretical prediction of the existence of pulsations in sdB stars,
due to an opacity bump associated with iron ionisation in
subphotospheric layers, was made by \citet{charpinet97}. Since both
$p$ and $g$-mode pulsations were discovered in sdB stars
\citep{kilkenny97,green03}, there has been a focus on the
possibilities to derive the internal structure and to put constraints
on the lesser known stages of the evolution by means of
asteroseismology.  Currently the immediate aims of asteroseismology of
sdB stars are to derive the mass of the Helium-burning core and
the mass of the thin Hydrogen envelope around the core
\citep[e.g.][]{randall06}, the rotational frequency and internal
rotation profile \citep{charpinet08},
the radius, and the composition of the core
\citep[e.g.][]{VanGrootel10,charpinet11b}.

Recent observational success has been achieved from splendid light
curves obtained by the CoRoT and \kep\ spacecrafts, delivering largely
uninterrupted time series with unprecedented accuracy for sdB stars.
Overviews of the \kep\ survey stage results for sdB stars were given
by \citet{ostensen10b,ostensen11b}, and case studies revealing dense
pulsational frequency spectra are presented by
\citet{reed10a} and \citet{baran11b}.
From \kep\ data it has become clear that the
$g$-modes in sdB stars can be identified from period spacings \citep{reed11c}.
Earth-size planets stripped from their outer layers have
been found around the pulsating sdB star KIC\,05807616
\citep[KOI-55, KPD\,1943+4058,][]{charpinet11b},
the star being also the subject of the first seismic study of an sdB star
in the \kep\ field \citep{VanGrootel10}.

\kep\ sdB+dM binaries with pulsating subdwarf components have been
presented by \citet{kawaler10b}, \citet{2m1938}, and by \citet{pablo11}.
White-dwarf companions in close \kep\ binaries are presented by
e.g.\ \citet{bloemen11,bloemen12} and \citet{silvotti12}.

\begin{figure*}[ht]
\centerline{\psfig{figure=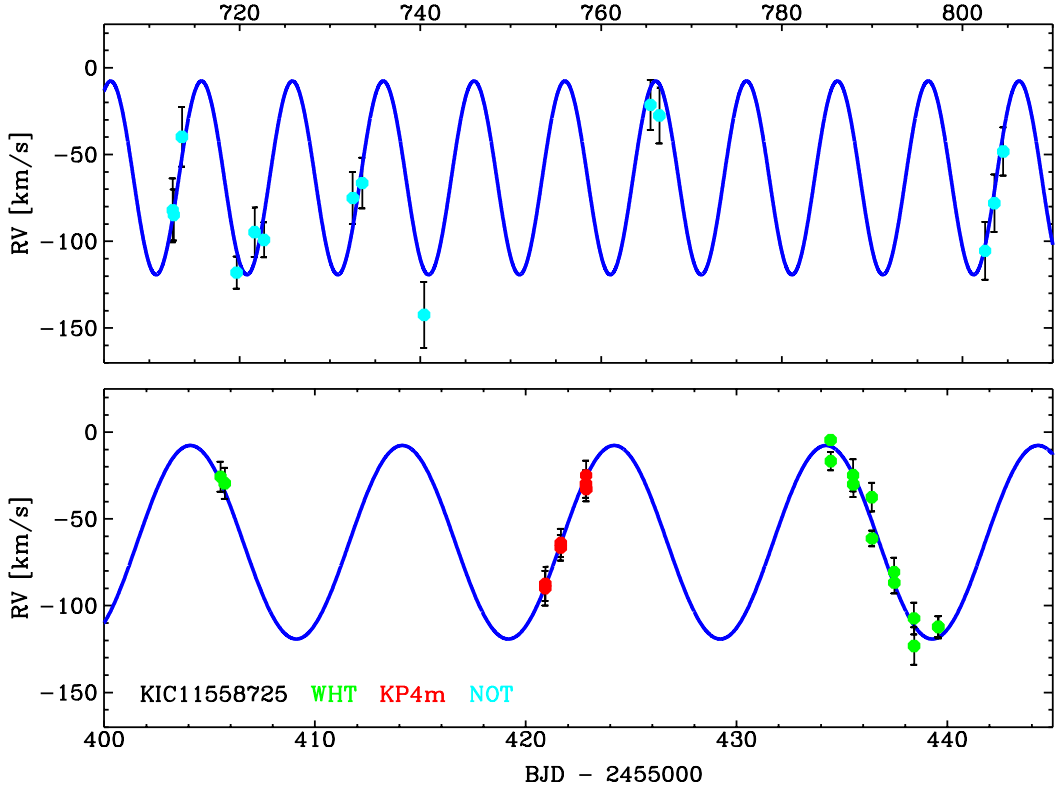,height=14cm,angle=-90}}
\caption[]{Radial-velocity curve from observations with the KP4m
  Mayall, the Nordic Optical, and the William Herschel telescopes. }
\label{fig:radialVelocityCurve}
\end{figure*}

Our target, \target\ or J19265+4930, is one of the 16 pulsating sdB
stars detected in the \kep\ field.  The \kep\ magnitude of \target\ is
 14.95, and the B-band magnitude is about 14.6, making it the
third brightest in the sample. A first description of the
spectroscopic properties, and the pulsational frequency spectrum as
found from the 26 day \kep\ survey dataset, was given by
\citet{ostensen11b}, with the source showing frequencies in the range
of 78--391\,$\mu$Hz.  Based on this relatively short data set already
36 frequencies were identified, showing the potential of this star for
a seismic study.  
Subsequently, \citet{baran11b} derived 53 frequencies in total from
the \kep\ survey data, and the frequencies were identified in terms of
spherical-harmonic degrees by \citet{reed11c}.
As a consequence, the star was  observed by
\kep\ from Q6 onwards.  At the time of writing, we have analysed
data from the five quarters Q6 to Q10.
We present the full frequency spectrum resulting from these 15 months
of short-cadence \kep\ observations.

In this paper we present our discovery of the binary nature of
\target\ based on low-resolution spectroscopy.  This object was
sampled as part of a spectroscopic observing campaign to study the
binary nature of the sdB population in the \kep\ field, for which
some preliminary results have already been presented by \citet{telting12a}.
Part of the data from this campaign were presented in case studies
on the close sdB+WD binary KIC\,06614501 \citep{silvotti12},
and the $p$-mode pulsating sdB KIC\,10139564 (Baran et al. 2012).

From our new spectra of \target\ we solve the orbital radial-velocity
amplitude, and derive a lower limit of the companion of the sdB star,
which is most likely an unseen white dwarf (Sect.\ 2).  
We use the average spectrum to study the atmospheric
parameters in detail.
We show that the orbital \kep\ light curve reveals strong evidence for
Doppler beaming that results in light variations at the 238\,ppm
level, consistent with theoretical predictions, again allowing us to
make an independent measurement of the orbital radial-velocity
amplitude (Sect.\ 3).
We extract 166 pulsational frequencies from the \kep\ light curve
(Sect.\ 4), and show that the orbit has an appreciable effect through
the light-travel time on the observed phases and frequencies of these
pulsations, which in fact allows us to make a third independent
measurement of the orbital-radial velocity amplitude (Sect.\ 5).
In the final sections of this paper we discuss pulsational period spacings
and frequency splittings, aiming to identify the spherical-harmonic
degree of the modes and to disclose the rotation period of the
subdwarf in \target.


\begin{table*}[ht]
\begin{center}
\caption[]{Log of the low-resolution spectroscopy of \target }
\label{tbl:obslog}
\begin{tabular}{ccrr@{~~}rll}
\hline\hline \noalign{\smallskip}
Mid-exposure Date    & {Barycentric JD} & S/N 
& RV~~~ & RV$_{\rm err}$~ & Telescope & Observer/PI \\
   & --2455000  & & km\,s$^{-1}$ & km\,s$^{-1}$ \\
\noalign{\smallskip} \hline \noalign{\smallskip}
2010-07-28 00:24:52.0 &  405.519273 &  113.9 &  $-$26.6 &   8.6 &  WHT  & JHT/CA \\
2010-07-28 05:16:04.8 &  405.721505 &   94.6 &  $-$30.6 &   9.0 &  WHT  & JHT/CA \\
2010-08-12 10:05:19.3 &  420.922376 &   57.6 &  $-$93.5 &  10.0 &  KP4m & MDR,LF \\
2010-08-12 10:16:03.9 &  420.929837 &   60.6 &  $-$91.1 &   9.8 &  KP4m & MDR,LF \\
2010-08-13 03:41:29.7 &  421.655828 &   68.3 &  $-$69.3 &   7.4 &  KP4m & MDR,LF \\
2010-08-13 03:52:04.9 &  421.663178 &   74.1 &  $-$66.4 &   8.1 &  KP4m & MDR,LF \\
2010-08-14 08:24:41.8 &  422.852489 &   59.1 &  $-$31.1 &   8.0 &  KP4m & MDR,LF \\
2010-08-14 08:35:42.3 &  422.860134 &   59.4 &  $-$25.9 &   8.5 &  KP4m & MDR,LF \\
2010-08-14 08:46:01.9 &  422.867305 &   60.1 &  $-$34.2 &   6.9 &  KP4m & MDR,LF \\
2010-08-25 22:54:52.3 &  434.456682 &   91.3 &   $-$4.6 &   2.1 &  WHT  & RH\O/CA \\
2010-08-25 23:09:46.0 &  434.467026 &  106.2 &  $-$17.4 &   5.2 &  WHT  & RH\O/CA \\
2010-08-27 00:36:26.9 &  435.527210 &  117.9 &  $-$25.8 &   9.2 &  WHT  & RH\O/CA \\
2010-08-27 00:46:36.6 &  435.534266 &  112.3 &  $-$31.2 &   7.3 &  WHT  & RH\O/CA \\
2010-08-27 21:56:54.7 &  436.416408 &   93.6 &  $-$39.0 &   8.2 &  WHT  & RH\O/CA \\
2010-08-27 22:08:29.4 &  436.424449 &   90.5 &  $-$63.8 &   4.4 &  WHT  & RH\O/CA \\
2010-08-28 23:18:34.2 &  437.473103 &   74.7 &  $-$83.9 &   8.3 &  WHT  & RH\O/CA \\
2010-08-28 23:28:43.8 &  437.480158 &   71.5 &  $-$90.5 &   6.1 &  WHT  & RH\O/CA \\
2010-08-29 21:53:35.3 &  438.414074 &  105.3 & $-$111.8 &   9.1 &  WHT  & RH\O/CA \\
2010-08-29 22:03:44.8 &  438.421129 &  107.7 & $-$128.2 &  10.9 &  WHT  & RH\O/CA \\
2010-08-31 01:31:47.4 &  439.565587 &  119.6 & $-$116.6 &   6.0 &  WHT  & RH\O/CA \\
2010-08-31 01:41:56.9 &  439.572642 &  109.0 & $-$117.0 &   6.4 &  WHT  & RH\O/CA \\
2011-05-31 01:56:05.9 &  712.581494 &   49.4 &  $-$85.7 &  18.4 &  NOT  & JHT \\
2011-05-31 04:48:22.9 &  712.701140 &   79.6 &  $-$88.5 &  14.7 &  NOT  & JHT \\
2011-06-01 02:35:14.4 &  713.608708 &   57.0 &  $-$41.4 &  17.1 &  NOT  & JHT \\
2011-06-07 03:46:43.9 &  719.658535 &   67.2 & $-$123.2 &   9.2 &  NOT  & JHT \\
2011-06-09 03:34:59.9 &  721.650444 &   49.6 &  $-$99.0 &  14.2 &  NOT  & JHT \\
2011-06-10 04:58:52.1 &  722.708717 &   43.9 & $-$103.3 &  10.0 &  NOT  & JHT \\
2011-06-20 00:20:10.7 &  732.515438 &   53.1 &  $-$78.3 &  14.9 &  NOT  & JHT \\
2011-06-21 01:27:07.4 &  733.561953 &   36.3 &  $-$69.3 &  14.7 &  NOT  & JHT \\
2011-06-27 21:53:09.0 &  740.413514 &   56.4 & $-$148.6 &  19.1 &  NOT  & JHT \\
2011-07-22 23:30:41.0 &  765.481614 &   38.1 &  $-$22.1 &  14.4 &  NOT  & JHT \\
2011-07-23 22:43:51.2 &  766.449100 &   49.2 &  $-$28.4 &  16.1 &  NOT  & JHT \\
2011-08-29 00:17:40.9 &  802.514159 &   42.1 & $-$110.0 &  16.6 &  NOT  & RO \\
2011-08-30 00:38:59.5 &  803.528944 &   48.1 &  $-$81.2 &  16.6 &  NOT  & RO \\
2011-08-31 00:25:05.0 &  804.519272 &   38.2 &  $-$50.1 &  13.9 &  NOT  & RO \\
\noalign{\smallskip} \hline
\end{tabular} \end{center} \end{table*}

\section{Spectroscopic observations}

Over the 2010 and 2011 observing seasons of the \kep\ field we
obtained altogether 35 spectra of \target. 

Low-resolution spectra (R\,$\approx$\,2000--2500) have been collected using
the Kitt Peak 4-m Mayall telescope with RC-Spec/F3KB, the kpc-22b
grating and a 1.5--2.0\,arcsec slit, the 2.56-m Nordic Optical Telescope
with ALFOSC, grism \#16 and a 0.5\,arcsec slit, and the 4.2-m William
Herschel Telescope with ISIS, the R600B grating and 0.8--1.0\,arcsec
slit.  Exposure times were 600\,s at KP4m and WHT, and either 600\,s
or 300\,s at the NOT.
The resulting resolutions based on the width of arc lines is 1.7\,\AA\ for
the KP4m and WHT setups, and 2.2\,\AA\ for the setup at the NOT. 
See Table~\ref{tbl:obslog} for an observing log.

The data were homogeneously reduced and analysed.  Standard reduction
steps within IRAF include bias subtraction, removal of pixel-to-pixel
sensitivity variations, optimal spectral extraction, and wavelength
calibration based on arc-lamp spectra. The target spectra and
the mid-exposure times were shifted to the barycentric frame of the solar
system.  The spectra were normalised to place the continuum at unity
by comparing with a model spectrum for a star with similar physical
parameters as we find for the target (see Sect.\ 3.2).
The mean spectra from each of the three telescopes are presented in Fig.\ 1.

\subsection{Radial velocities and orbit solution}

Radial velocities were derived with the FXCOR package in IRAF. We used
the H$\gamma$, H$\delta$, H$\zeta$ and H$\eta$ lines to determine the
radial velocities (RVs), and used the spectral model fit (see next
section) as a template.  See Table~\ref{tbl:obslog} for the results,
with errors in the radial velocities as reported by FXCOR.  The errors
reported by FXCOR are correct relative to each other, but may need
scaling depending on, amongst other things, the parameter settings and
the validity of the template as a model of the star.  As our fit
results in a $\chi^2$-value close to unity (see Table 2) we trust that
scaling of the FXCOR errors is not necessary and that the derived
errors of the fit parameters are adequate.

Assuming a circular orbit we find an orbital period of 10.0545(48)\,d,
with a radial-velocity amplitude of 58.1(1.7)\,km\,s$^{-1}$ for the
subdwarf. See Table~\ref{tbl:orbit} for the complete parameter
listing.  The radial velocities and the derived solution are shown in
Fig.\ 2.  When fitting an eccentric radial-velocity curve the
amplitude goes down to 56.8\,km\,s$^{-1}$, for eccentricity
$e$=0.063(33).  Throughout this paper we regard the orbit as circular.

Given the solution presented in Table~\ref{tbl:orbit}, the orbital
radius of the sdB star can be approximated by $a_{\rm sdB} \sin
i$\,=\,11.5\,$R_{\sun}$, which corresponds to a light-travel time of
53.6(1.6) seconds between orbital phases corresponding to closest and
furthest distance to the Sun.  

The orbital solution combined with the mass function gives a lower
limit for the mass of the companion of more than 0.63\,$M_{\sun}$, if
one assumes a canonical mass of 0.48\,$M_{\sun}$ for the sdB star.  As
the spectrum does not reveal clear evidence for light contribution
from a companion, it should be either an unseen compact
object, or an unseen K star. 
As the 2MASS J\,=\,15.38(5) and H\,=\,15.35(9) magnitudes do not indicate
a rising infrared flux that would reveal the presence of a K star in this
system, we conclude that the unseen companion is most likely a
white dwarf.
If the inclination of the system is smaller than $i$\,$\la$\,40 degrees then
the companion may be a neutron star or black hole ($i$\,$\la$\,25 degrees).

Assuming a radius for the sdB star of 0.2\,$R_{\sun}$, { and
assuming a white dwarf companion, one may expect} to see eclipses
only if the inclination angle is higher than $i$\,$\ga$\,88 degrees.
{ We did not detect eclipses (see Fig.\ 3).}



\begin{table}[b]
\begin{center}
\caption[]{Orbital solution of \target}
\label{tbl:orbit}
\begin{tabular}{lr@{~}l}
\hline \hline \noalign{\smallskip}
 system velocity [km\,s$^{-1}$]              & $-$66.1   &  (1.4)  \\
 radial-velocity amplitude $K$ [km\,s$^{-1}$]&  58.1     &  (1.7)  \\
 period $P$ [day]                    &  10.0545  &  (0.0048) \\
 phase [BJD $-$ 2455000]             & 421.682   &  (0.062)  \\
 $\chi^2$                            &  34.48                \\
 reduced $\chi^2$                    &   1.11                \\
\noalign{\smallskip} \hline
\end{tabular} \end{center} \end{table}

\subsection{Atmospheric parameters}

The spectra were shifted to remove the orbital motion, before being
co-added to obtain high-S/N spectra (KP4m S/N\,=\,175; NOT
S/N\,=\,145; WHT S/N\,=\,305) with minimal orbital line broadening,
for all three observatories.  We derive the atmospheric parameters of
the star from each of these mean spectra, and produce a weighted
mean using the formal fitting errors as variance weights, as listed in
Table~\ref{tbl:specpar}.  Our final adopted values are
\teff\,=\,27910(210)\,K, \logg\,=\,5.410(15)\,dex and
\lheh\,=\,$-$3.116(18)\,dex, which are quite compatible with
the parameters \teff\,=\,27400\,K, \logg\,=\,5.37, \lheh\,=\,$-$2.8
found from the initial survey spectrum in \citet{ostensen11b}.
{ The errors on the adopted values are the errors on the weighted
  means, and reflect the spread of the individual measurements
  rather than the formal errors.  Systematic errors related to model
  physics typically are of the order 500\,K, 0.05, and 0.05, for 
\teff, \logg, and \lheh, respectively.}

\begin{table}[b]
\begin{center}
\caption[]{Atmospheric parameters of the subdwarf in \target}
\label{tbl:specpar}
\begin{tabular}{llll}
\hline \hline \noalign{\smallskip}
Telescope   & \teff      & \logg & \lheh \\
            &          K &  cm\,s$^{-2}$ &       \\
\noalign{\smallskip} \hline \noalign{\smallskip}
 KP4m       & 27950(110) & 5.378(20) & --3.078(42) \\
 NOT        & 27410(110) & 5.418(20) & --3.124(32) \\
 WHT        & 28010(50)  & 5.417(11) & --3.125(26) \\
\noalign{\smallskip} \hline \noalign{\smallskip}
 adopted    & 27910(210) & 5.410(15) & --3.116(18) \\
\noalign{\smallskip} \hline
\end{tabular} \end{center} \end{table}

\begin{figure*}[t]
\centerline{\psfig{figure=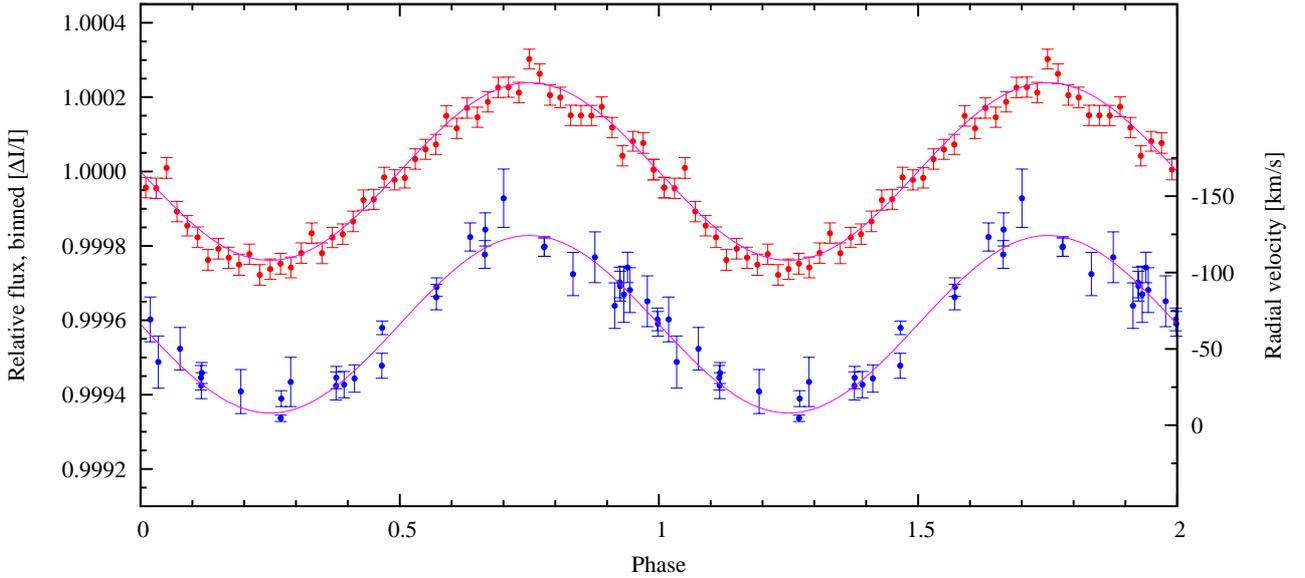,height=8cm}}
\caption[]{Top: the 461 day long \kep\ light curve from quarters Q6 through Q10, 
  folded on the orbital period and binned into 50 bins. Bottom: phased
  radial velocities from the spectra.  }

\label{fig:keplerOrbit}
\end{figure*}

Besides the lines of the hydrogen Balmer series and \ion{He}{i} lines,
clear lines of magnesium (4481\,\AA), and silicon (4552\AA, 4567\AA)
are present. The high S/N mean WHT spectrum seen in the upper panel of
Fig.~\ref{fig:meanSpectra} also allows the identification of a number
of lines from \ion{O}{ii}, \ion{N}{ii} as well as the \ion{C}{ii}
doublet at 4267\AA\ and the \ion{C}{iii} lines at 4647 and 4650\AA. By
comparing the strength of these lines relative to LTE model spectra
with explicit metal lines computed for the adopted parameters listed
in Table~\ref{tbl:specpar}, varying only the metallicity relative to
solar, we find that oxygen is depleted 
to 4\% of the solar abundance, nitrogen to 13\%, and carbon to 1.6\%.
Such high depletion is normal in sdB stars due to gravitational
settling, and large deviations from the solar mix for individual
elements are also common \citep{heber00}.

\section{The orbital light curve from \kep\ photometry}

We analysed the \kep\ light curve of \target\ as obtained in quarters
Q6--Q10, totalling roughly 15 months of data with 58.8\,s sampling
time \citep[ i.e. short-cadence data; see][]{gilliland10b}. The
photometric time series spans BJD 2455372 to 2455833, which is roughly
the same range as our spectroscopic dataset.  We analyse the light
curve as transformed to fractional intensities $\Delta$I/I.

At first we used the { raw} light curves produced by the \kep\ data
processing pipeline \citep[SAP\_FLUX;][]{jenkins10}.
These light curves are extracted using { a standard small pixel mask}
selected to minimise the noise in the output data. However, during the
observations slight drifts were induced by guiding corrections and
thermal effects in the focal plane, which implies that some light
drifts in or out of the {  small pixel mask}, causing trends that must
be removed. For pulsations with much shorter timescales than the
trends, this is not a problem, but with a 10\,d orbital period in
\target, our concern was that the detrending could suppress the
orbital beaming effect.  We therefore analysed also the raw
\kep\ pixel data using {  a large custom pixel mask including all 20 to
  23 pixels with target signal, as opposed to 4 to 6 pixels in the
  small pixel mask of the standard pipeline extraction.}  This
recovers 36\%\ more flux from the target and the monthly chunks can be
detrended with simple linear trends.  This approach gave a beaming
amplitude that is 34\%\ higher than our first estimate from the
pipeline-reduced data.  However, these large-aperture light curves are
substantially more noisy and contain many more bad points. Hence we
only use the {  raw, custom-mask} pixel data to determine the amplitude
of the Doppler beaming, but use the { standard-mask pipeline-reduced
  raw} data for our pulsation-frequency analysis in Sect.\ 4.

A Fourier analysis of the \kep\ light curve reveals the orbital period
to be 10.0516(27)\,d, in perfect agreement with the period found from
spectroscopy.  In Fig.~\ref{fig:keplerOrbit} we show the \kep\ light
curve folded on the orbital period into 50 phase bins.  A sine fit to
the folded light curve, leaving only the amplitude as a free
parameter, gives an amplitude of the orbital effects seen by \kep\ of
238(6) ppm, and this amplitude is consistent with that derived from
the Fourier analysis.  The fit is shown as a solid line in the figure.
We do not find orbital harmonics.

\subsection{Doppler beaming}

The high precision \kep\ data permits us to accurately explore the
low-amplitude Doppler beaming effect, something that is very hard to do with
ground based data. This effect is induced by stellar motion in a binary orbit
and causes brightness modulation \citep{RL79}.
The Doppler beaming effect permits an estimate of the radial velocity
without resorting to spectroscopic data. This effect was detected for the
first time in \kep\ data by \citet{VanKerkwijk10} and for a planetary system
in {\em CoRoT} data by \citet{MF10}.
A confirmation of the correspondence between Doppler beaming amplitudes and
radial velocities was first established by \citet{bloemen11}.

The light curve of \target\ displays a brightening of the sdB star at
the orbital phases where the star is approaching us in its orbit
(i.e.~when its { orbital} radial velocities are negative, see
Fig.~\ref{fig:keplerOrbit}), and this is exactly the effect expected
by Doppler beaming. Note that unlike \object{KPD\,1946+4330}
\citep{bloemen11}, \target\ does not show any sign of ellipsoidal
deformation which in the closest sdB+WD binaries produces a strong
harmonic signal at \porb/2 \citep[see also e.g.][]{silvotti12}.  This
is consistent with the much longer orbital period of \target\ as
opposed to the 0.4\,d period of \object{KPD\,1946+4330}.

\begin{figure*}[t]
\centerline{\psfig{figure=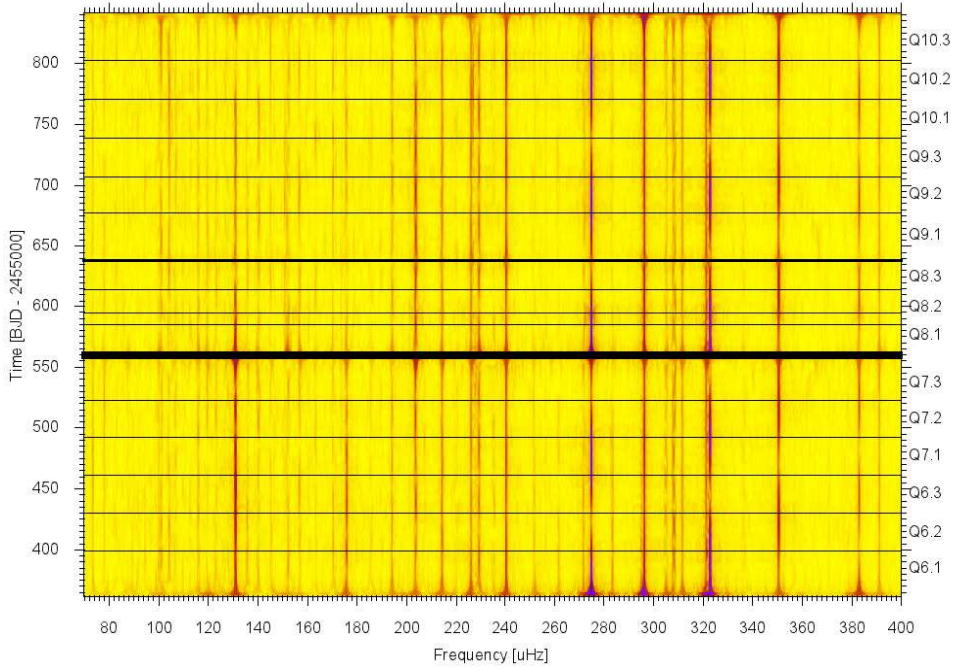,height=9cm}}
\caption[]{A dynamic Fourier spectrum from quarters Q6 through Q10
  computed for 20 day stretches of data.  The pronounced beatings show up as
  resolved frequencies in the Fourier spectrum of the complete data
  set. The horizontal lines reflect data gaps. } 
\label{fig:runningFT}
\end{figure*}

In the case of Doppler beaming the observed flux from the target,
$F_\lambda$, is related to the emitted spectrum and orbital velocity as
\begin{equation}
F_\lambda = F_{0,\lambda} \left( 1 - B {v_r\over c} \right)
\end{equation}
and the beaming amplitude relates to the orbital radial-velocity
amplitude $K$ as 
\begin{equation}
A_B ~ D = B {K\over c} ~ ,
\end{equation}
where $ A_B $ is the amplitude of the beaming signal in the light curve
(see Fig. 3), and $D$ the decontamination factor discussed below.

The beaming factor $B$ contains three terms. For an approaching
source a contribution of +1 comes from the enhanced photon arrival
rate, and another term of +2 from the geometrical aberration of the
wavefronts.  The last term comes from the Doppler shift of the
spectrum, which for a hot sdB star produces a negative effect as the
blue-dominated flux gets blue-shifted out of the observed
\kep\ bandpass when the star is approaching.  We compute this last
term using TMAP model spectra \citep{werner03a} that cover the entire
\kep\ bandpass, following the procedure described in
\citet{bloemen11}.

With the beaming factor $B$=\,1.403(5), and the spectroscopically
derived orbital radial-velocity amplitude, we compute a predicted
total beaming amplitude for the \kep\ bandpass of 270\,ppm.  The
observed beaming amplitude of $A_B$=238\,ppm is 88\%\ of the predicted
value.

The \kep\ pixels onto which our target is imaged suffer from
contamination from neighbouring objects, and from passing charge from
brighter sources when clocking out the CCD.  According to the
\kep\ Target
catalogue\footnote{http://archive.stsci.edu/kepler/kepler\_fov/search.php},
the contamination factors for \target\ are 0.098 (Q6,Q10), 0.110 (Q7),
0.079 (Q8) and 0.106 (Q9), or 0.0982 on average, implying that all
periodic amplitudes derived from these \kep\ fluxes should be
multiplied by $D$\,=\,1/(1$-$0.0982)\,=\,1.109 to get the intrinsic
amplitudes of \target.  When applying this decontamination factor we
find that the photometric amplitude of the Doppler beaming as seen by
\kep\ is consistent with the spectroscopically derived radial-velocity
amplitude within the errors of the data.  Given the observed amplitude
$A_B$, we use Eq.\ (2) to derive a value of
$K$=56.5(1.4)\,km\,s$^{-1}$ for the orbital radial-velocity amplitude,
from Doppler beaming.

The fact that this value is consistent with that of the spectroscopic
value proves that the companion of the sdB does not significantly
{ contribute to} the observed Doppler beaming, consistent with a compact
nature of the companion.


\section{The pulsation spectrum}

We are interested in the pulsation frequencies in \target\ for three
reasons.  Firstly, we intend to use the pulsations as clocks to derive
the orbital light-travel time delay and consequently another
independent determination of the orbital radial-velocity amplitude.
Secondly, the pulsations may reveal the internal structure of the
subdwarf through a detailed seismic study, for which we intend to
derive mode identifications in Sect.\ 6.  Furthermore, the
frequency splittings may disclose the rotation period of the subdwarf.

To investigate the pulsational frequency spectrum we used the
\kep\ light curves as produced by the standard \kep\ data processing
pipeline.  The fractional object intensities $\Delta$I/I of quarters
Q6--Q10 were first detrended, and subsequently outlying points were
discarded leaving 625617 data points with standard deviation
$\sigma$=2600ppm\,.  An iterative { prewhitening} process,
involving a standard fast-fourier transform to find peaks and
subsequent non-linear least-squares (NLLS) sine-curve fits to subtract
the pulsational content mode by mode from the original data, was used
to derive the frequency list of Table 4.  In total we extracted 166
pulsational frequencies, in addition to the orbital frequency.  The
standard deviation of the data after removal of all significant
periodic signals was $\sigma$=1900ppm\,.

The standard deviation in the Fourier amplitude spectrum of the
original data amounts to $\sigma_{\rm FT}$=4.2\,ppm, and
we adopt 4$\sigma_{\rm FT}$=17\,ppm as the threshold of significance
of the peaks in the Fourier amplitude spectrum.

We note that the observed amplitudes in Table 4 have not been
corrected for the \kep\ decontamination factor $D$ (see above), nor for the
amplitude smearing due to the effective exposure time of 58.8 seconds.
Both effects cause the observed amplitudes to be smaller than the
intrinsic amplitudes, and the latter affects mostly the shortest
periods ($p$-modes).  For the strongest pulsation, with a period of
3641.1\,s, the exposure time leads to a decrease of the amplitude of
4.3\%, while for the shortest extracted period of 197.9\,s the amplitude
decrease is 14\%.

To illustrate the fact that all pulsational frequencies are present
throughout the \kep\ run, we show a section of the Fourier transform
in a dynamic form in Fig.\ 4.  There is clear beating among the stronger
frequencies; in fact, these beatings show up as resolved frequencies in the
Fourier transform of the full \kep\ data set, and may be attributed
to the rotation of the subdwarf (Sect.\ 7).

\subsection{Combination frequencies}

We do not find evidence for harmonics of the pulsation frequencies.
We do, however, find evidence for combination frequencies.  When
computing the residual frequency $\delta f = f_3 - f_2 - f_1$ for all
combinations of 168 frequencies in Table 4, we find 13 combinations
that are consistent with a residual $\delta f$=0 within the errors of
the frequencies.  
Thirty-five different frequencies are involved in these 13 combinations.
One of the combinations is an orbital alias discussed in
Sect.\ 5.1.  See Table 5 for the list of frequency combinations.

To investigate the significance of this, we take our observed
frequencies and perturb each of them by a random offset which is big
enough to upset real combinations, but small enough such that the
distribution of these randomised frequencies looks like that of the observed
distribution. We perturb the frequencies by $\pm$1\,$\mu$Hz.  For each
randomised frequency we adopt the frequency error of the original
unperturbed peak.  For 50 sets of such randomised frequencies, we find
an average of 3.7 combinations that satisfy $\delta f$=0 within the
errors of the frequencies.

Hence, for the set of 13 combinations that we find in our 168
frequencies, we expect that about 4 are by-chance combinations.  We
assume that the combinations with the lowest-amplitude peaks are most
likely to be by-chance combinations, as these peaks have the largest
error on their frequency. These probable frequency combinations are
marked in Table 4, and we take care when assigning these modes in period
spacings and frequency multiplets.

\begin{table*}[ht]
\begin{center}

\caption[]{Extracted pulsational frequencies. }

\label{tbl:freqlist}
\begin{tabular}{rrrrrccl}
\hline\hline \noalign{\smallskip}
\multicolumn{1}{c}{Frequency}&  \multicolumn{1}{c}{Period}    & Amplitude  &  S/N  &  Splitting  & $n_{\ell=1}$ & $n_{\ell=2}$  &   \\
\multicolumn{1}{c}{$\mu$Hz}  &   \multicolumn{1}{c}{s}   & \multicolumn{1}{r}{ppm} &    &  \multicolumn{1}{r}{$\mu$Hz} \\
\noalign{\smallskip} \hline \noalign{\smallskip}
5053.4659\,(20) &   197.88399\,(0.00008) &    23\,(4) &   5.4 &        &      &      &       \\
3103.8490\,(17) &   322.18062\,(0.00017) &    28\,(4) &   6.5 &  1.58  &      &      &       \\
3102.2724\,(16) &   322.34436\,(0.00016) &    29\,(4) &   6.8 &        &      &      &       \\
3075.0214\,(09) &   325.20099\,(0.00010) &    50\,(4) &  11.7 &  1.55  &      &      &       \\
3073.4671\,(07) &   325.36545\,(0.00008) &    63\,(4) &  14.8 &  1.15  &      &      &       \\ 
3072.3162\.(24) &   325.48733\,(0.00026) &    19\,(4) &   4.4 &        &      &      &  OA   \\
3059.8960\,(08) &   326.80849\,(0.00008) &    59\,(4) &  13.8 &        &      &      &       \\
1463.3977\,(25) &   683.34123\,(0.00116) &    18\,(4) &   4.2 &        &      &  -1  &       \\
1444.0731\,(19) &   692.48572\,(0.00091) &    24\,(4) &   5.6 &        &      &      &  *    \\
1422.9117\,(11) &   702.78430\,(0.00053) &    43\,(4) &  10.1 &        &      &      &       \\
1420.4369\,(17) &   704.00875\,(0.00083) &    27\,(4) &   6.3 &  0.24  &      &      &       \\
1420.1919\,(22) &   704.13022\,(0.00111) &    20\,(4) &   4.7 &        &      &      &  *    \\
1399.2750\,(23) &   714.65579\,(0.00119) &    20\,(4) &   4.7 &        &      &      &       \\
1396.5427\,(21) &   716.05400\,(0.00109) &    22\,(4) &   5.1 &  0.13  &      &      &       \\
1396.4129\,(20) &   716.12059\,(0.00105) &    23\,(4) &   5.4 &  0.27  &      &      &       \\
1396.1397\,(20) &   716.26072\,(0.00104) &    23\,(4) &   5.4 &        &      &      &  *    \\
1370.4145\,(11) &   729.70621\,(0.00057) &    43\,(4) &  10.1 &        &      &      &  *    \\
1308.1017\,(23) &   764.46658\,(0.00136) &    20\,(4) &   4.7 &        &      &      &  *    \\
1292.1409\,(13) &   773.90943\,(0.00077) &    35\,(4) &   8.2 &        &      &      &  *    \\
1270.2485\,(13) &   787.24755\,(0.00083) &    34\,(4) &   8.0 &        &      &      &       \\
1265.4160\,(19) &   790.25394\,(0.00116) &    25\,(4) &   5.8 &        &      &      &       \\
1237.3675\,(20) &   808.16736\,(0.00134) &    22\,(4) &   5.1 &        &      &   0  &       \\
1139.1458\,(23) &   877.85072\,(0.00175) &    20\,(4) &   4.7 &        &      &      &       \\
1128.9600\,(21) &   885.77098\,(0.00162) &    22\,(4) &   5.1 &        &      &      &       \\
1083.6435\,(26) &   922.81273\,(0.00220) &    18\,(4) &   4.2 &        &   -2 &      &  *    \\
1042.2709\,(20) &   959.44349\,(0.00188) &    22\,(4) &   5.1 &  0.53  &      &  \ua &       \\
1041.7417\,(12) &   959.93086\,(0.00115) &    37\,(4) &   8.7 &  0.26  &      &      &       \\
1041.4816\,(20) &   960.17062\,(0.00183) &    23\,(4) &   5.4 &  0.27  &      &   1  &       \\
1041.2151\,(14) &   960.41633\,(0.00131) &    32\,(4) &   7.5 &  0.52  &      &      &       \\
1040.6927\,(11) &   960.89840\,(0.00104) &    41\,(4) &   9.6 &  0.54  &      &      &       \\
1040.1546\,(20) &   961.39554\,(0.00187) &    23\,(4) &   5.4 &  0.58  &      &      &       \\
1039.5769\,(08) &   961.92981\,(0.00073) &    58\,(4) &  13.6 &        &      &  \da &       \\
 915.9922\,(20) &  1091.71231\,(0.00240) &    23\,(4) &   5.4 &        &      &   2  &       \\
 888.0072\,(20) &  1126.11695\,(0.00248) &    23\,(4) &   5.4 &        &      &      &       \\
 875.0949\,(18) &  1142.73315\,(0.00237) &    25\,(4) &   5.8 &        &   -1 &      &  *    \\
 826.2266\,(13) &  1210.32167\,(0.00189) &    35\,(4) &   8.2 &        &      &      &       \\
 804.6305\,(09) &  1242.80646\,(0.00134) &    53\,(4) &  12.4 &  0.44  &      &  \ua &       \\
 804.1940\,(26) &  1243.48108\,(0.00408) &    17\,(4) &   4.0 &  1.07  &      &   3  &       \\
 803.1235\,(14) &  1245.13844\,(0.00219) &    32\,(4) &   7.5 &  0.56  &      &      &       \\
 802.5678\,(17) &  1246.00067\,(0.00267) &    27\,(4) &   6.3 &        &      &  \da &       \\
 774.4270\,(13) &  1291.27732\,(0.00217) &    35\,(4) &   8.2 &        &      &      &       \\
 741.5079\,(26) &  1348.60322\,(0.00468) &    18\,(4) &   4.2 &        &      &      &       \\
 729.2202\,(24) &  1371.32792\,(0.00447) &    19\,(4) &   4.4 &        &      &      &       \\
 712.5805\,(03) &  1403.35031\,(0.00067) &   135\,(4) &  31.7 &  0.40  &  \ua &  \ua &  *    \\
 712.1802\,(07) &  1404.13908\,(0.00137) &    66\,(4) &  15.5 &  0.41  &      &      &       \\
 711.7661\,(05) &  1404.95599\,(0.00098) &    92\,(4) &  21.6 &  0.37  &      &   4  &       \\
 711.3935\,(05) &  1405.69176\,(0.00104) &    87\,(4) &  20.4 &  0.34  &    0 &      &       \\
 711.0567\,(09) &  1406.35756\,(0.00183) &    49\,(4) &  11.5 &        &  \da &  \da &       \\
 685.7967\,(20) &  1458.15803\,(0.00431) &    23\,(4) &   5.4 &        &      &      &       \\
 683.9097\,(23) &  1462.18144\,(0.00502) &    19\,(4) &   4.4 &        &      &      &       \\
 663.5667\,(11) &  1507.00754\,(0.00253) &    41\,(4) &   9.6 &  0.52  &      &      &       \\
 663.0469\,(19) &  1508.18887\,(0.00439) &    24\,(4) &   5.6 &        &      &      &       \\
 659.0236\,(15) &  1517.39631\,(0.00335) &    32\,(4) &   7.5 &  0.34  &      &  \ua &       \\
 658.6855\,(14) &  1518.17521\,(0.00322) &    33\,(4) &   7.7 &  0.44  &      &      &       \\
 658.2465\,(15) &  1519.18768\,(0.00341) &    31\,(4) &   7.2 &  0.42  &      &   5  &       \\
 657.8284\,(06) &  1520.15327\,(0.00139) &    76\,(4) &  17.8 &        &      &  \da &  *    \\
 639.4081\,(06) &  1563.94636\,(0.00144) &    78\,(4) &  18.3 &  0.48  &      &      &  *    \\
 638.9238\,(22) &  1565.13179\,(0.00543) &    21\,(4) &   4.9 &  0.36  &      &      &  *    \\
 638.5688\,(11) &  1566.00196\,(0.00265) &    42\,(4) &   9.8 &  0.42  &      &      &       \\
 638.1471\,(18) &  1567.03676\,(0.00446) &    25\,(4) &   5.8 &        &      &      &       \\
 603.9510\,(24) &  1655.76352\,(0.00670) &    19\,(4) &   4.4 &        &    1 &      &       \\
 595.5174\,(18) &  1679.21205\,(0.00494) &    26\,(4) &   6.1 &        &      &   6  &       \\
 581.4725\,(25) &  1719.77176\,(0.00730) &    19\,(4) &   4.4 &        &      &      &       \\
\noalign{\smallskip} \hline
\end{tabular} \end{center} \end{table*}

\addtocounter{table}{-1}
\begin{table*}[ht]
\begin{center}
\caption[]{Extracted pulsational frequencies, continued.}
\begin{tabular}{rrrrrccl}
\hline\hline \noalign{\smallskip}
\multicolumn{1}{c}{Frequency}&  \multicolumn{1}{c}{Period}    & Amplitude  &  S/N  &  Splitting  & $n_{\ell=1}$ & $n_{\ell=2}$  &   \\
\multicolumn{1}{c}{$\mu$Hz}  &   \multicolumn{1}{c}{s}   & \multicolumn{1}{r}{ppm} &    &  \multicolumn{1}{r}{$\mu$Hz} \\
\noalign{\smallskip} \hline \noalign{\smallskip}
 559.9286\,(26) &  1785.94207\,(0.00833) &    17\,(4) &   4.0 &        &      &  \ua &       \\
 557.6564\,(21) &  1793.21886\,(0.00676) &    22\,(4) &   5.1 &  1.38  &      &   7  &       \\
 556.2798\,(13) &  1797.65640\,(0.00413) &    36\,(4) &   8.4 &        &      &  \da &       \\
 536.5423\,(09) &  1863.78608\,(0.00318) &    50\,(4) &  11.7 &  0.48  &      &      &       \\
 536.0656\,(20) &  1865.44339\,(0.00686) &    23\,(4) &   5.4 &        &      &      &       \\
 511.0034\,(07) &  1956.93427\,(0.00272) &    64\,(4) &  15.0 &        &      &   8  &  *    \\
 456.4978\,(21) &  2190.59091\,(0.01022) &    21\,(4) &   4.9 &        &      &      &       \\
 435.3696\,(18) &  2296.89908\,(0.00970) &    25\,(4) &   5.8 &        &      &      &  *    \\
 390.8675\,(02) &  2558.41205\,(0.00129) &   233\,(4) &  54.8 &  0.29  &      & \ua  &  *    \\
 390.5727\,(20) &  2560.34269\,(0.01316) &    23\,(4) &   5.4 &        &      &  12  &       \\
 382.7335\,(02) &  2612.78405\,(0.00124) &   253\,(4) &  59.5 &  0.14  &  \ua &      &       \\
 382.5913\,(01) &  2613.75508\,(0.00076) &   416\,(4) &  97.8 &        &    5 &      &       \\
 370.4606\,(12) &  2699.34236\,(0.00879) &    38\,(4) &   8.9 &        &      &  13  &       \\
 351.9691\,(10) &  2841.15865\,(0.00788) &    47\,(4) &  11.0 &        &      &      &       \\
 350.2917\,(02) &  2854.76388\,(0.00125) &   299\,(4) &  70.3 &  0.12  &  \ua & \ua  &       \\
 350.1676\,(01) &  2855.77522\,(0.00056) &   668\,(4) & 157.1 &        &    6 &  14  &  *    \\
 336.8906\,(17) &  2968.32256\,(0.01499) &    27\,(4) &   6.3 &  0.34  &      & \ua  &  *    \\
 336.5482\,(22) &  2971.34244\,(0.01924) &    21\,(4) &   4.9 &  0.17  &      &      &       \\
 336.3773\,(15) &  2972.85207\,(0.01342) &    30\,(4) &   7.0 &  0.40  &      &  15  &       \\
 335.9730\,(12) &  2976.43006\,(0.01099) &    37\,(4) &   8.7 &  0.40  &      &      &       \\
 335.5761\,(14) &  2979.94986\,(0.01211) &    34\,(4) &   8.0 &        &      & \da  &       \\
 322.5146\,(01) &  3100.63481\,(0.00057) &   779\,(4) & 183.2 &  0.12  &  \ua &      &       \\
 322.3960\,(01) &  3101.77527\,(0.00071) &   620\,(4) & 145.8 &  0.61  &    7 &      &       \\
 321.7834\,(05) &  3107.68011\,(0.00502) &    88\,(4) &  20.7 &  0.43  &      & \ua  &       \\
 321.3523\,(06) &  3111.84932\,(0.00603) &    74\,(4) &  17.4 &  0.42  &      &  16  &  OA   \\
 320.9336\,(01) &  3115.90941\,(0.00083) &   537\,(4) & 126.3 &        &  \da & \da  &       \\
 311.3506\,(01) &  3211.81286\,(0.00140) &   339\,(4) &  79.7 &        &      &      &       \\
 308.5923\,(08) &  3240.52146\,(0.00817) &    59\,(4) &  13.8 &  0.16  &      & \ua  &       \\
 308.4370\,(05) &  3242.15289\,(0.00510) &    95\,(4) &  22.3 &  0.34  &      &      &       \\
 308.0959\,(03) &  3245.74214\,(0.00263) &   184\,(4) &  43.2 &  0.43  &      &  17  &       \\
 307.6612\,(02) &  3250.32863\,(0.00219) &   220\,(4) &  51.7 &        &      & \da  &  *    \\
 305.0273\,(02) &  3278.39459\,(0.00262) &   188\,(4) &  44.2 &  0.42  &      &      &       \\
 304.6094\,(07) &  3282.89325\,(0.00777) &    63\,(4) &  14.8 &  0.40  &      &      &       \\
 304.2116\,(08) &  3287.18530\,(0.00845) &    58\,(4) &  13.6 &        &      &      &  *    \\
 296.1067\,(01) &  3377.16086\,(0.00066) &   796\,(4) & 187.2 &  0.92  &  \ua & \ua  &       \\
 295.1847\,(05) &  3387.70993\,(0.00587) &    89\,(4) &  20.9 &        &    8 &  18  &       \\
 283.2231\,(05) &  3530.78551\,(0.00580) &    98\,(4) &  23.0 &  0.32  &      & \ua  &       \\
 282.9004\,(11) &  3534.81263\,(0.01348) &    42\,(4) &   9.8 &  0.36  &      &  19  &       \\
 282.5393\,(09) &  3539.33115\,(0.01066) &    54\,(4) &  12.7 &        &      & \da  &       \\
 274.7524\,(01) &  3639.64034\,(0.00100) &   611\,(4) & 143.7 &  0.11  &  \ua &      &  *    \\
 274.6415\,(01) &  3641.11018\,(0.00068) &   891\,(4) & 209.6 &        &    9 &      &       \\
 271.9863\,(05) &  3676.65540\,(0.00682) &    91\,(4) &  21.4 &  0.14  &      & \ua  &       \\
 271.8489\,(14) &  3678.51341\,(0.01850) &    34\,(4) &   8.0 &  0.14  &      &  20  &  *    \\
 271.7083\,(06) &  3680.41699\,(0.00748) &    83\,(4) &  19.5 &  0.38  &      &      &       \\
 271.3237\,(06) &  3685.63425\,(0.00758) &    82\,(4) &  19.2 &        &      & \da  &       \\
 256.2548\,(08) &  3902.36612\,(0.01218) &    58\,(4) &  13.6 &  0.12  &  \ua &      &       \\
 256.1353\,(23) &  3904.18665\,(0.03544) &    20\,(4) &   4.7 &        &   10 &      &       \\
 254.0371\,(15) &  3936.43229\,(0.02272) &    31\,(4) &   7.2 &        &      &      &       \\
 240.4970\,(04) &  4158.05561\,(0.00728) &   109\,(4) &  25.6 &  0.14  &  \ua &      &       \\
 240.3552\,(01) &  4160.50933\,(0.00144) &   557\,(4) & 131.0 &        &   11 &      &       \\
 235.6888\,(09) &  4242.88224\,(0.01617) &    51\,(4) &  12.0 &        &      &  24  &  *    \\
 229.4898\,(01) &  4357.49261\,(0.00270) &   325\,(4) &  76.4 &  0.13  &      &      &       \\
 229.3626\,(07) &  4359.90906\,(0.01269) &    69\,(4) &  16.2 &  0.13  &      &      &       \\
 229.2375\,(02) &  4362.28846\,(0.00335) &   261\,(4) &  61.4 &  1.24  &      &      &       \\
 227.9950\,(08) &  4386.06201\,(0.01570) &    56\,(4) &  13.1 &        &  \ua & \ua  &       \\
 226.1883\,(01) &  4421.09470\,(0.00206) &   434\,(4) & 102.1 &        &   12 &  25  &       \\
 214.8602\,(05) &  4654.18897\,(0.01078) &    92\,(4) &  21.6 &  0.41  &  \ua & \ua  &       \\
 214.4538\,(01) &  4663.00855\,(0.00262) &   381\,(4) &  89.6 &        &   13 &  27  &       \\
 204.0365\,(01) &  4901.08376\,(0.00331) &   333\,(4) &  78.3 &  0.31  &  \ua &      &       \\
 203.7217\,(01) &  4908.65758\,(0.00262) &   420\,(4) &  98.8 &        &   14 &      &       \\
 194.4829\,(04) &  5141.84071\,(0.01050) &   115\,(4) &  27.0 &  0.25  &  \ua &  \ua &       \\
 194.2320\,(02) &  5148.48327\,(0.00583) &   208\,(4) &  48.9 &        &   15 &  30  &       \\
 185.9787\,(12) &  5376.95929\,(0.03514) &    38\,(4) &   8.9 &  0.23  &  \ua &      &       \\
\noalign{\smallskip} \hline                                              
\end{tabular} \end{center} \end{table*}

\addtocounter{table}{-1}
\begin{table*}[ht]
\begin{center}
\caption[]{Extracted pulsational frequencies, continued.}
\begin{tabular}{rrrrrccl}
\hline\hline \noalign{\smallskip}
\multicolumn{1}{c}{Frequency}&  \multicolumn{1}{c}{Period}    & Amplitude  &  S/N  &  Splitting  & $n_{\ell=1}$ & $n_{\ell=2}$  &   \\
\multicolumn{1}{c}{$\mu$Hz}  &   \multicolumn{1}{c}{s}   & \multicolumn{1}{r}{ppm} &    &  \multicolumn{1}{r}{$\mu$Hz} \\
\noalign{\smallskip} \hline \noalign{\smallskip}
 185.7520\,(09) &  5383.52087\,(0.02650) &    50\,(4) &  11.7 &        &   16 &      &  *    \\
 183.1588\,(16) &  5459.74210\,(0.04837) &    28\,(4) &   6.5 &        &      &      &       \\
 173.8372\,(09) &  5752.50813\,(0.03087) &    49\,(4) &  11.5 &        &      &      &       \\
 170.5770\,(05) &  5862.45349\,(0.01651) &    95\,(4) &  22.3 &  0.27  &  \ua &  \ua &       \\
 170.3090\,(06) &  5871.67950\,(0.02219) &    71\,(4) &  16.7 &        &   18 &  35  &  *    \\
 163.4371\,(06) &  6118.56357\,(0.02383) &    72\,(4) &  16.9 &        &   19 &  37  &       \\
 160.8351\,(13) &  6217.54915\,(0.05208) &    34\,(4) &   8.0 &        &      &      &  *    \\
 156.9593\,(03) &  6371.07649\,(0.01259) &   149\,(4) &  35.0 &  0.10  &  \ua &      &       \\
 156.8617\,(07) &  6375.04157\,(0.02766) &    68\,(4) &  16.0 &        &   20 &      &       \\
 152.6142\,(04) &  6552.46996\,(0.01839) &   107\,(4) &  25.1 &  0.69  &      &  \ua &       \\
 151.9237\,(05) &  6582.25067\,(0.02006) &    99\,(4) &  23.2 &  0.72  &      &  40  &       \\
 151.2009\,(10) &  6613.71905\,(0.04546) &    46\,(4) &  10.8 &  0.08  &  \ua &      &       \\
 151.1162\,(09) &  6617.42475\,(0.03923) &    53\,(4) &  12.4 &  0.22  &   21 &      &       \\
 150.8946\,(10) &  6627.14155\,(0.04458) &    45\,(4) &  10.5 &        &  \da &      &       \\
 146.8171\,(11) &  6811.19663\,(0.05080) &    42\,(4) &   9.8 &        &      &  42  &       \\
 140.2879\,(05) &  7128.19949\,(0.02788) &    83\,(4) &  19.5 &        &   23 &  44  &       \\
 135.9729\,(05) &  7354.40412\,(0.02817) &    88\,(4) &  20.7 &  0.30  &  \ua &      &  *    \\
 135.6687\,(11) &  7370.89720\,(0.06169) &    40\,(4) &   9.4 &        &   24 &      &       \\
 131.1578\,(01) &  7624.40245\,(0.00545) &   486\,(4) & 114.3 &  0.37  &  \ua &      &  *    \\
 130.7852\,(04) &  7646.12630\,(0.02057) &   130\,(4) &  30.5 &        &   25 &      &       \\
 123.4258\,(10) &  8102.03660\,(0.06511) &    46\,(4) &  10.8 &  0.38  &  \ua & \ua  &       \\
 123.0479\,(07) &  8126.91403\,(0.04350) &    69\,(4) &  16.2 &        &   27 &  51  &       \\
 119.8705\,(04) &  8342.33836\,(0.02940) &   108\,(4) &  25.4 &        &   28 &      &       \\
 116.1148\,(03) &  8612.16711\,(0.02395) &   142\,(4) &  33.4 &        &   29 &      &  *    \\
 104.3568\,(03) &  9582.50490\,(0.02625) &   160\,(4) &  37.6 &  0.36  &  \ua &  61  &       \\
 104.0008\,(07) &  9615.31517\,(0.06067) &    70\,(4) &  16.4 &        &   33 &      &  *    \\
 101.5403\,(05) &  9848.30600\,(0.05298) &    84\,(4) &  19.7 &  0.41  &  \ua &  \ua &  *    \\
 101.1341\,(05) &  9887.85717\,(0.04481) &   100\,(4) &  23.5 &  0.59  &   34 &  63  &       \\
 100.5486\,(15) &  9945.44211\,(0.14413) &    31\,(4) &   7.2 &  0.37  &      &      &       \\
 100.1774\,(05) &  9982.29229\,(0.05201) &    88\,(4) &  20.7 &        &      &  64  &       \\
  94.7358\,(12) & 10555.67247\,(0.13476) &    38\,(4) &   8.9 &        &      &  68  &       \\
  92.2056\,(19) & 10845.32955\,(0.22353) &    24\,(4) &   5.6 &  0.47  &      &      &       \\
  91.7333\,(14) & 10901.16609\,(0.16491) &    33\,(4) &   7.7 &        &      &  70  &       \\
  89.4365\,(15) & 11181.11603\,(0.19143) &    30\,(4) &   7.0 &  1.56  &      &      &       \\
  87.8741\,(11) & 11379.91791\,(0.14794) &    40\,(4) &   9.4 &        &      &      &       \\
  86.0158\,(16) & 11625.76501\,(0.21525) &    29\,(4) &   6.8 &        &      &      &       \\
  83.2845\,(13) & 12007.03178\,(0.19272) &    34\,(4) &   8.0 &        &      &  78  &       \\
  78.2703\,(04) & 12776.24508\,(0.06990) &   107\,(4) &  25.1 &        &      &      &       \\
  73.6556\,(04) & 13576.69795\,(0.08256) &   102\,(4) &  24.0 &        &   49 &  89  &  *    \\
  49.7780\,(11) & 20089.18155\,(0.45588) &    40\,(4) &   9.4 &        &      &      &  *    \\
  37.8572\,(18) & 26415.06723\,(1.22748) &    26\,(4) &   6.1 &        &      &      &       \\
\noalign{\smallskip} \hline                                              
\end{tabular} \end{center} 
\tablefoot{
The columns marked $n_{\ell}$ list the 
sequence number obtained from period spacing, for the $\ell$=1 and $\ell$=2 sequences.  
Multiplet candidates used to derive the period spacings are indicated by the span of
the arrows above and below the sequence numbers; for doublets only an up-arrow is 
used.  The last column lists OA for orbital alias, or an
asterisk if the frequency is part of a combination, as listed in Table 5. 
}
\end{table*}

\section{Using the pulsations as clocks to derive the orbit}

In analogy to the effect that the Earth's orbit has on arrival times
of a variational signal, and on the apparent frequencies of that signal,
the orbit of the sdB alters the phases and frequencies of its
pulsations as perceived by a distant observer.  

As we already established in Sect.\ 2, the light-travel time between
conjunctions is 53.6\,s, which amounts to a significant phase change
for the 300 second $p$-mode pulsations in this star.  Similarly, the
Doppler frequency change of the pulsations at orbital quadratures
amounts to $\Delta f$ = $f$ $*$ $K / c$, which, with $K$=58\,km\,s$^{-1}$,
even for $g$-modes gives frequency shifts of 0.02--0.2\,$\mu$Hz. These
are certainly measurable with our 460 day data set with frequency
resolution of 0.025$\mu$Hz.  If the orbit is known, one can simply
correct for these effects by converting all observation timings to the
centre-of-mass frame of the system, equivalent to the common approach
of conversion of observation timings to the barycentric frame of the
solar system.

If the orbit is not known, one can use the pulsations as
clocks 
{ \citep[see e.g.][]{hulse75}}
to derive the light-travel time that corresponds to the radius
of the orbit, which relates to the radial-velocity amplitude $K$
that can be obtained from spectroscopy or the Doppler-beaming curve as
\begin{eqnarray}
\Delta t_{\rm R} = \frac{a_{\rm sdB} \sin i }{c}= \frac{K}{c} \frac{P_{\rm orb}}{2\pi}
\sqrt{1-e^2} ~ ,
\end{eqnarray}
where we introduce the R\o mer delay, $\Delta t_{\rm R}$, to represent
the light-travel time.

For a circular orbit, the light-travel delay as a function of the
subdwarf's position in its orbit can be written as
\begin{eqnarray}
T_{\rm delay}(t) =  \Delta t_{\rm R} ~
\cos\left(\frac{2\pi}{P_{\rm orb}} (t - T_{\rm orb})\right) ~ ,
\end{eqnarray}
where $T_{\rm orb}$ is the time at which the subdwarf is closest to
the Sun in its orbit, corresponding to the orbital phase listed in Table 2.

We did not detect any second- or third-harmonic peaks, neither for the
orbit nor for the pulsations.  Hence, for sinusoidal signals the
\kep\ light curve of \target\ can be approximated by

\begin{eqnarray}
\frac{\Delta I(t)}{I} & = &  A_B  \sin \left( \frac{2\pi}{P_{\rm orb}}(t - T_{\rm
  orb}) \right)  \nonumber \\
 & + &  \sum_{i} A_{\rm i,puls}  \sin \left( 2\pi F_{\rm i,puls} 
(t - T_{\rm i,puls} + T_{\rm delay}(t) ) \right)   ~ ,
\end{eqnarray}
where the first term describes the orbital beaming effect, and where
all individual pulsations are affected by the same orbital
light-travel delay $T_{\rm delay}(t)$.  
{ Here, the phase of the individual pulsations, $T_{\rm i,puls}$, is expressed
in the time domain rather than as an angle.}
Note that the above sum of sine
curves is equivalent to the model that we fit as part of the
prewhitening procedure described in Sect.\ 4, with the addition of a
phase delay that introduces just one extra parameter, i.e. the
amplitude $\Delta t_{\rm R}$ of the light-travel delay.

We apply the above Eq.\ (5) to derive a third independent
measurement of the radial-velocity amplitude of the subdwarf in
\target.  For this purpose, we included the above phase delay in our
NLLS Levenberg-Marquardt routine, and fitted the above model (5) to the
\kep\ Q6--Q10 light curve.  In the model we included the 70 strongest
frequencies that are identified to be part of $\ell$=1 and $\ell$=2
sequences of the subdwarf (see next section), as to avoid possible
pulsations from the unseen component in the binary.

Simultaneously fitting all 70 pulsational amplitudes, frequencies and phases,
together with the R\o mer delay and phase as free parameters, we find
$\Delta t_{\rm R}$= 26.5(1.5)\,s and $T_{\rm orb}$= 421.744(88)\,[BJD
 -- 2455000], which is in perfect agreement with the spectroscopic
parameters in Table 2, and the Doppler beaming amplitude.  Given the
observed R\o mer delay $\Delta t_{\rm R}$, we derive a value of
$K$=57.5(3.2)\,km\,s$^{-1}$ for the orbital radial-velocity amplitude, from
the light-travel time effect.

\begin{figure*}[t]
\centering
\psfig{figure=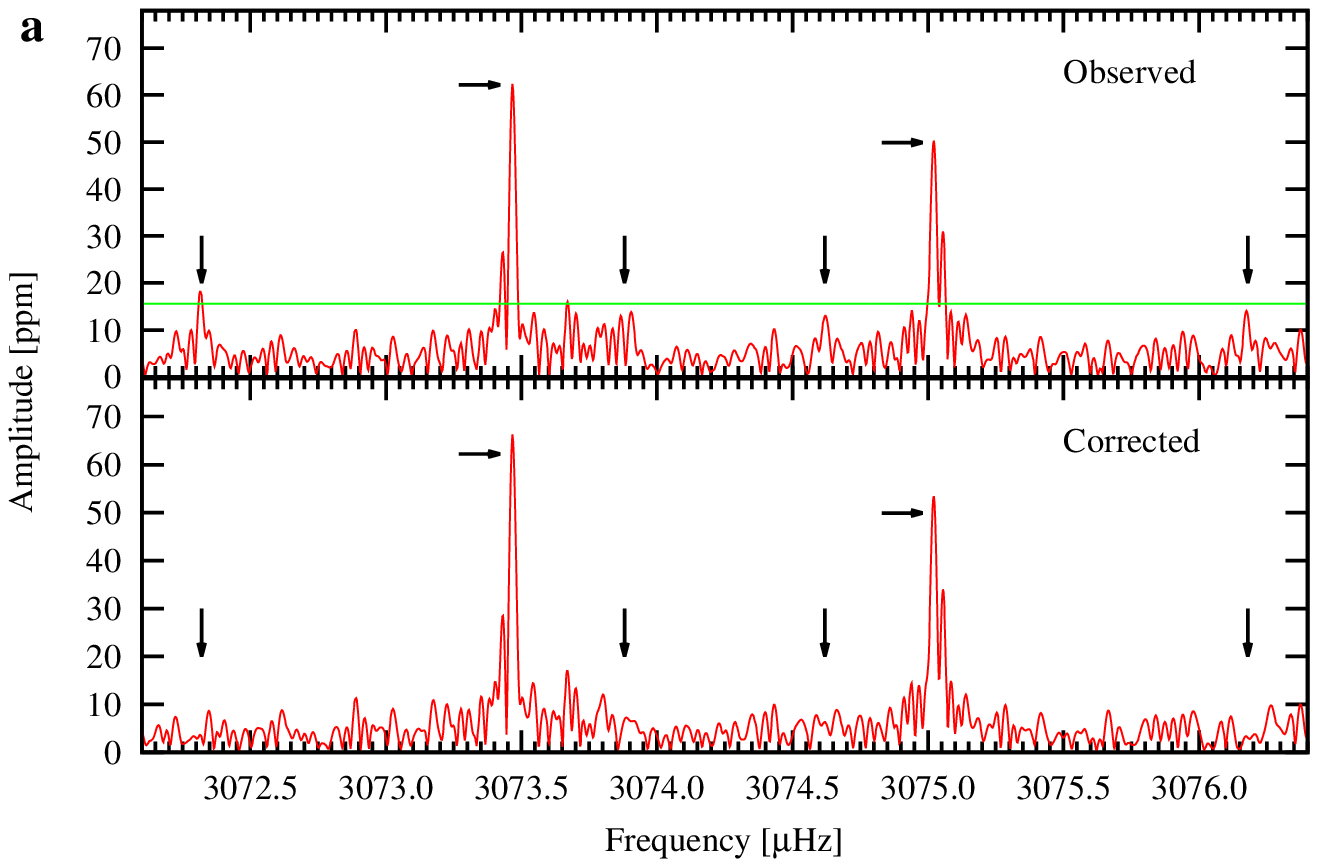,angle=270,width=9cm}
\psfig{figure=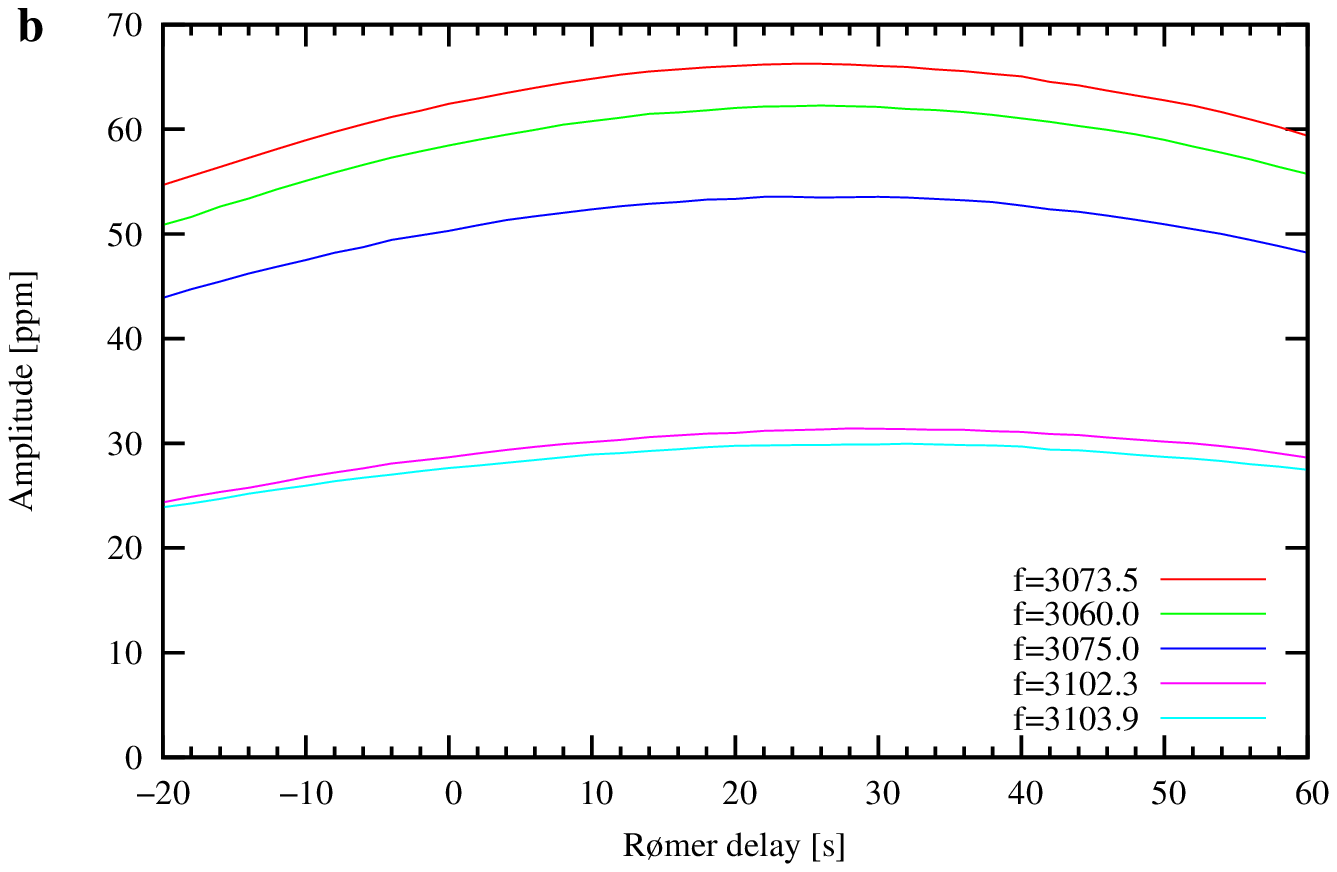,angle=270,width=9cm}\\
\psfig{figure=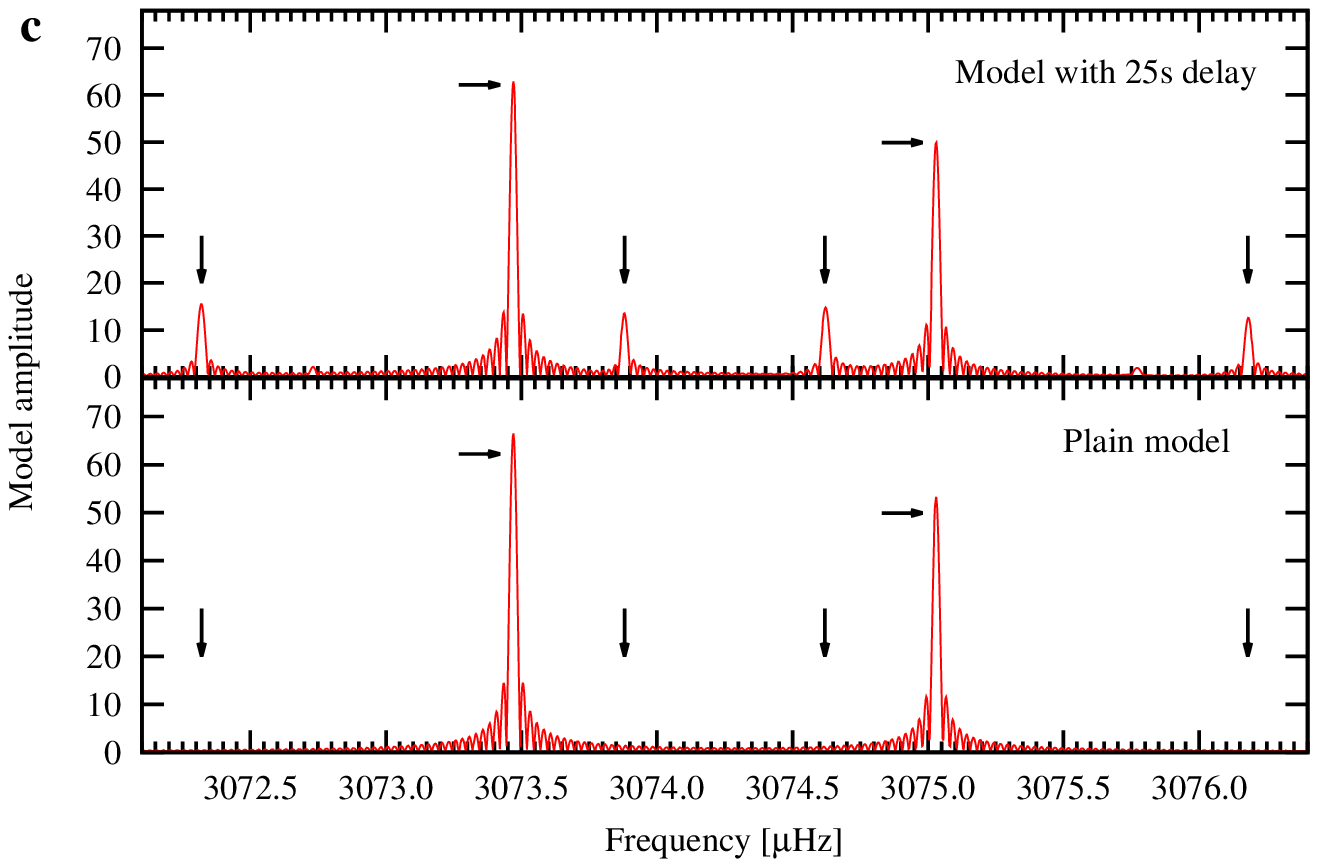,angle=270,width=9cm}
\psfig{figure=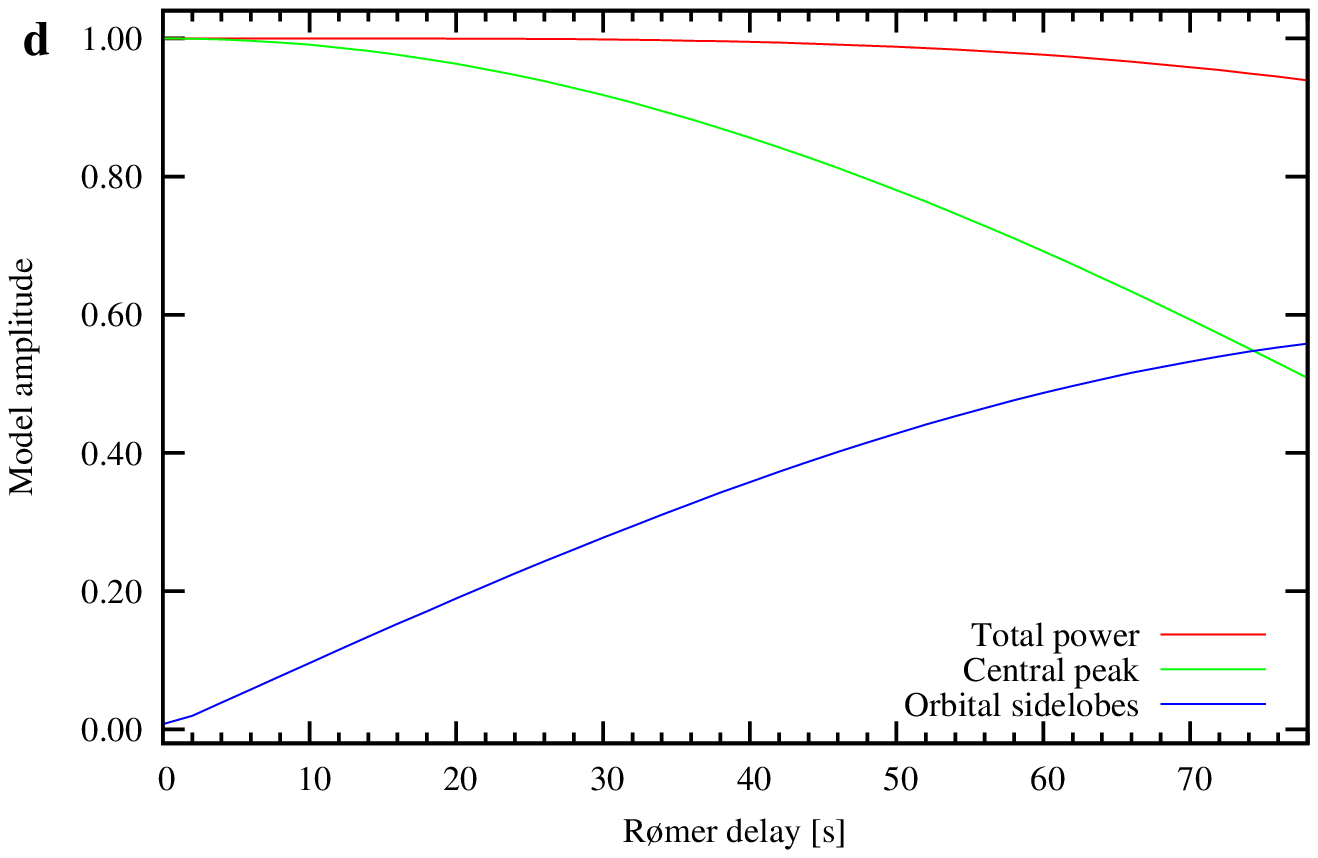,angle=270,width=9cm}
\caption[]{a: The pulsation spectrum \target\ as observed (upper panel), and
after correcting for the R\o mer delay (lower panel). The line in the upper panel
indicates our 4-$\sigma$ detection limit.
b: The amplitude of five pulsation peaks as a function of R{\o}mer correction.
c: Same as a), but for a model light curve with the same duration and sampling
frequency as the data in a).
d: The predicted amplitude of the central peak and first alias
as computed for a pure sine function distorted by an increasing
R{\o}mer delay. The upper curve indicates the total power of the
central peak and the aliases. This sum departs from unity when the
unaccounted-for second orbital alias pair starts to contribute. {
  Entries in the legends have the same top-to-bottom order as the curves.}}
\label{fig:romerdelay}
\end{figure*}

\begin{table*}
\begin{center}
\caption{Possible frequency combinations}
\label{tbl:combinationfreqs}
\begin{tabular}{rrrrrrr}
\hline \hline \noalign{\smallskip}
$F_1$~ ~ ~ ~ & $F_2$~ ~ ~ ~ &  $F_3$ ~ ~ ~ ~ &$F_3-F_2-F_1$& $A_1$&$A_2$& $A_3$\\
$\mu$Hz ~ ~ ~ ~ & $\mu$Hz~ ~ ~ ~ &  $\mu$Hz ~ ~ ~ ~ & $\mu$Hz ~ ~ ~ & ppm &ppm& ppm \\
\noalign{\smallskip} \hline \noalign{\smallskip}
   1.1521\,(02)& 3072.3162\,(24) & 3073.4671\,(07) &-0.0012 ~ ~ & 223 &  19 & 63 \\
  49.7780\,(11)&  135.9729\,(05) &  185.7520\,(09) & 0.0011 ~ ~ &  40 &  88 & 50 \\
  49.7780\,(11)& 1370.4145\,(11) & 1420.1919\,(22) &-0.0006 ~ ~ &  40 &  43 & 20 \\
  73.6556\,(04)&  638.9238\,(22) &  712.5805\,(03) & 0.0011 ~ ~ & 102 &  21 &135 \\
 101.5403\,(05)&  170.3090\,(06) &  271.8489\,(14) &-0.0004 ~ ~ &  84 &  71 & 34 \\
 104.0008\,(07)& 1292.1409\,(13) & 1396.1397\,(20) &-0.0020 ~ ~ &  70 &  35 & 23 \\
 116.1148\,(03)&  274.7524\,(01) &  390.8675\,(02) & 0.0003 ~ ~ & 142 & 611 &233 \\
 131.1578\,(01)&  304.2116\,(08) &  435.3696\,(18) & 0.0002 ~ ~ & 486 &  58 & 25 \\
 135.9729\,(05)& 1308.1017\,(23) & 1444.0731\,(19) &-0.0015 ~ ~ &  88 &  20 & 24 \\
 160.8351\,(13)&  350.1676\,(01) &  511.0034\,(07) & 0.0007 ~ ~ &  34 & 668 & 64 \\
 235.6888\,(09)&  639.4081\,(06) &  875.0949\,(18) &-0.0020 ~ ~ &  51 &  78 & 25 \\
 307.6612\,(02)&  350.1676\,(01) &  657.8284\,(06) &-0.0004 ~ ~ & 220 & 668 & 76 \\
 336.5482\,(22)& 1083.6435\,(26) & 1420.1919\,(22) & 0.0002 ~ ~ &  21 &  18 & 20 \\
\noalign{\smallskip} \hline
\end{tabular}
\end{center} 
\tablefoot{
 All combinations $F_3$\,=\,$F_2+F_1$ are significant to
  within the errors of the observed frequencies. The rightmost columns
  list the amplitudes. The  top entry is an orbital alias (see Sect.\ 5.1)}
\end{table*}

\subsection{Orbital aliases in the Fourier domain}

Above we have solved the orbit by directly fitting the light curve
with a parameterised model of the pulsations that includes the orbital
light-travel delay. A different approach was recently investigated by
\citet{SK12}, who demonstrated how one can derive the orbit from
the ratio of the amplitude of the central peak to the orbital
aliases that the R\o mer delay introduces in the observed Fourier
spectrum. 

In the case of \target\ we find that, of our 166 extracted pulsation
frequencies, only a { few of the strongest \mbox{g-modes} and a} few of
the highest-frequency pulsation modes in the weak $p$-mode regime have
orbital aliases strong enough to be close to the amplitude level of
significance.

{ Around the strongest g-modes the orbital sidelobes appear in
  frequency regions populated by weak unresolved structure and
  prewhitening residuals.  A more clear example of the orbital sidelobes}
  is given by the two peaks at 3073.5 and 3075.0 $\mu$Hz, shown in the
  top panel of Fig.~5a.  The sidelobes of each of the two central
  peaks can just be discerned close to the 4.0-$\sigma$ line at a
  separation of $F_{\rm orb}$\,=\,1.15\,$\mu$Hz. 
  Since the amplitudes of these aliases are { hardly significant,
   direct application of the ratio method of \citet{SK12} leads to
   inaccurate results for this particular dataset.}

The two central peaks, however, are at $\sim$16\,$\sigma$, and their
amplitudes can be measured with very high precision. We can therefore
exploit the energy theorem in Fourier analysis that implies that the
total power in the Fourier transform should not change due to the
time-distorting orbital motion. All the power observed in the
sidelobes will therefore be restored into the central peak if the
light curve is corrected for the light-travel time. In the bottom half
of Fig.~\ref{fig:romerdelay}a we show that this is indeed the
case. When we perform a light-travel time correction on the
observation timings of the light curve, in the same way as the
barycentric correction is done to remove the spacecraft orbit, then
the orbital aliases in the Fourier domain disappear and the power is
transferred to the central peak.  For comparison,
Fig.~\ref{fig:romerdelay}c shows a simulation using noise-free,
evenly-sampled data modelled with and without an imposed R\o mer
delay.

\begin{table}[b]
\begin{center}
\caption[]{Measured R\o mer delay from high-frequency
  pulsations. }
\label{tbl:romer}
\begin{tabular}{lll}
\hline \hline \noalign{\smallskip}
Frequency  & \multicolumn{1}{c}{$\Delta t_R$} & \multicolumn{1}{c}{$A_0$} \\
~ ~ ~  $\mu$Hz  & \multicolumn{1}{c}{s} & \multicolumn{1}{c}{ppm} \\
\noalign{\smallskip} \hline \noalign{\smallskip}
3059.896 & 26.04(05) & 62.2(1) \\
3073.467 & 25.45(05) & 66.2(1) \\
3075.021 & 26.25(07) & 53.6(1) \\
3102.271 & 29.35(15) & 31.4(2) \\
3103.849 & 29.23(13) & 29.9(1) \\
\noalign{\smallskip} \hline
\end{tabular} \end{center} 
\tablefoot{
The amplitude $A_0$ is the recovered amplitude when applying the
  corresponding R\o mer delay $\Delta t_R$.
}
\end{table}

We can also use the amplitude of the central peak to measure the
R\o mer delay, and get another estimate of the radial-velocity
amplitude.  If we take our observed light curve and adjust the
observation timings for various values of the R\o mer delay,
we find that the Fourier amplitudes of the pulsation peaks
vary as shown in Fig.~\ref{fig:romerdelay}b. Clearly, all five modes
reach maximum amplitude for roughly the same delay correction.  The
delays and peak amplitudes are given in Table~\ref{tbl:romer}, and if
we take an average using the amplitudes as weights we find $\Delta
t_R$\,=\,26.4(0.7)\,s. The gain in amplitude for these modes is about 6\%,
when the optimal R\o mer delay is applied.

\begin{figure*}[t]
\centering
\psfig{figure=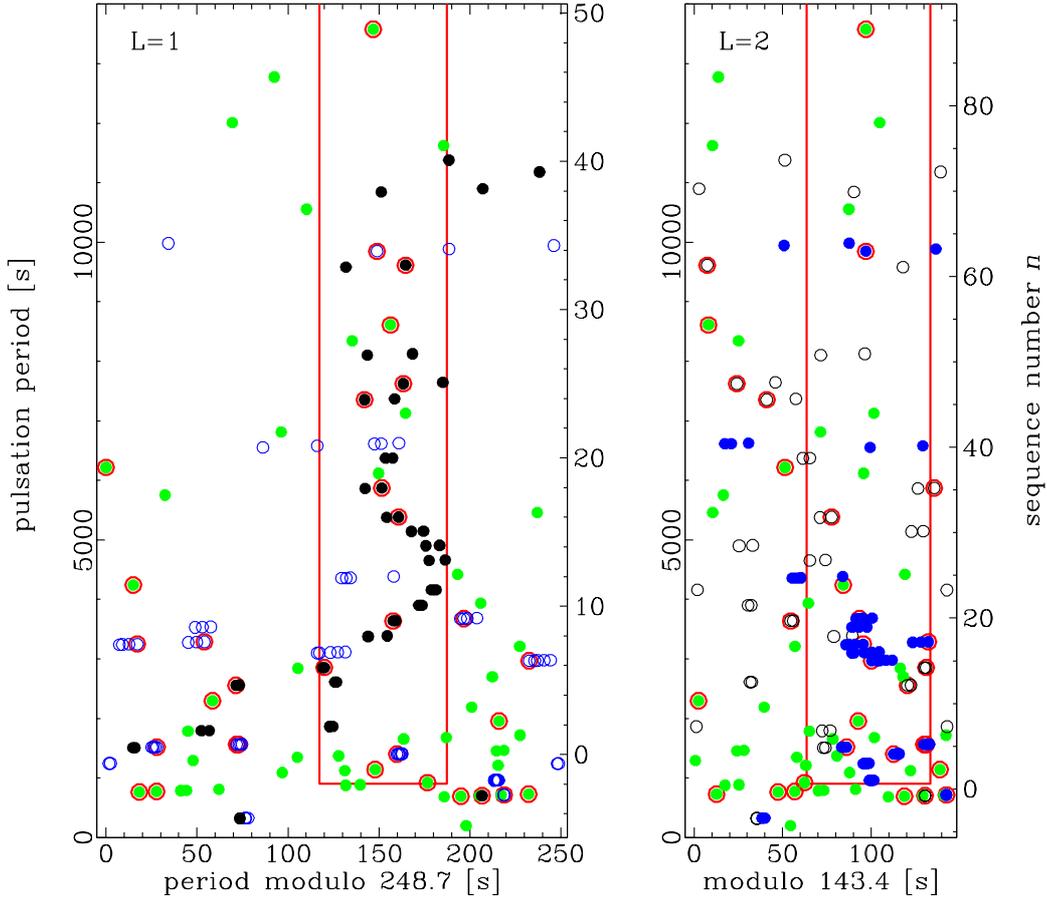,width=14cm}
\caption[]{
{ Echelle diagrams of the $\ell$\,=\,1 and $\ell$\,=\,2 period-spacing
  sequences. The red rectangles depict the areas of acceptance to
  these sequences.  Single frequencies are plotted in light green,
  doublets in black, and multiplets in blue (see Table 4).  Modes
  marked with red circles form frequency combinations (Table 5).  }}
\label{fig:spacings}
\end{figure*}

A model of how the observed amplitude of a pulsation peak taken
to reside at 3073.5\,$\mu$Hz\ drops with increasing $\Delta t_R$ is shown
in Fig.~\ref{fig:romerdelay}d. The corrected amplitude of the 
central peak is found to be 6.1\%\ higher in a simulated light curve
with no R\o mer delay than in one with a 26\,s delay, as observed.
The predicted amplitude of the orbital alias is found to be 25.2\%.
Rearranging Eq.~(27) in \citet{SK12} in terms of this ratio, we find
\begin{equation}
{{A_{+1} + A_{-1}}\over{2 A_0}} = {{\pi \Delta t_R} \over {P_{\rm osc}}} ~ ,
\end{equation}
which corresponds to 24.2\%\ for $\Delta t_R$\,=\,25\,s and $P_{\rm
  osc}$\,=\,325\,s, which demonstrates that our numerical modelling
(Fig. 5d) is consistent with their analytical derivation.  

In Fig.~5d we show that the total power contained in the central peak
and its aliases is preserved while applying observation-timing
corrections.  The curve drops off for large delays as we have only
accounted for the first pair of a full comb of orbital aliases.


\section{Pulsational period spacings}

Recently, \citet{reed11c} have revealed that $g$-modes in sdB stars
show sequences with almost fixed period spacings.  This is contrary to
what models predicted up to now, i.e. strongly variable period
spacings caused by mode trapping due to a highly stratified internal
structure.  The observed regular period spacings indicate that
internal mixing processes must be considerably stronger than
presumed. As the period spacings follow the asymptotic
relation for $g$-modes, they change with the value of the
spherical-harmonic degree $\ell$ of the modes. Hence, the spacings
allow us to determine the $\ell$-value of the modes directly.

\citet{reed11c} already identified $\ell$ values for the
pulsations that could be resolved in the \kep\ survey data of
\target.  They identified an $\ell$=1 sequence with period spacing
$\Delta \Pi_{\ell=1}$=246.8\,s, and an $\ell$=2 sequence with period
spacing $\Delta \Pi_{\ell=2}$=142.6\,s.

Here we use the same methods of \citet{reed11c} to derive the period
spacings, using the much longer Q6--Q10 data set.  First we defined
the period spacing for $\ell$=1, as marked in Table 4.  This was
achieved iteratively.  As $\ell$=1 modes suffer the least geometric
cancellation \citep[see e.g.][]{reed05}, and hence should have highest
amplitudes if all modes are driven at a similar level, we started off
with the highest-amplitude modes.  Subsequently we worked with an
$\ell$=1 sequence consisting only of doublets, and finally all the
$\ell$=1 periods marked in Table 4. For each step we derived mutually
consistent values of the period spacing. We recover all the $\ell$=1
identifications from \citet{reed11c} except for two periods not
detected in the Q6--Q10 data.

Including all 57 $\ell$=1 multiplet candidate peaks from Table 4 in a
linear regression $P(n)$\,=\,$C$\,+\,$n*\Delta \Pi_{\ell=1}$, we find
$\Delta \Pi_{\ell=1}$=248.87(18)\,s with RMS=19\,s.  Here, the mode
sequence number $n$ should be equal to the radial number of nodes in
the pulsation, $n'$, but shifted by a fixed offset that we cannot
determine.  When fitting for each of the 29 multiplets (mostly
doublets) only the peak closest to this first fit, we obtain a
consistent period spacing $\Delta\Pi_{\ell=1}$=248.68(23)\,s, with
RMS=14\,s.

Given that for the asymptotic $g$-mode domain we expect $\Delta
\Pi_{\ell=2}$=$\Delta \Pi_{\ell=1} /\sqrt 3$, we used a spacing of
$\Delta \Pi_{\ell=2}$=143\,s as a starting point for the $\ell$=2
sequence.  Again we iterated adding the highest-amplitude peaks and
some frequency multiplets that cannot be $\ell$=1 modes.  Including
all 76 $\ell$=2 multiplet candidate peaks in a linear regression, we
find $\Delta \Pi_{\ell=2}$=143.44(08)\,s with RMS=16\,s.  When fitting
for each of the 34 multiplets only the peak closest to this initial
fit, we obtain a consistent period spacing
$\Delta\Pi_{\ell=2}$=143.37(10)\,s, with RMS=14\,s.

Of 166 pulsation frequencies, 23 have periods shorter than 900\,s
which we did not include as $g$-mode period-spacing candidates.  From
the 143 $g$-mode pulsations we find that 108 can be matched with
either the $\ell$=1 or $\ell$=2 sequence, when accepting periods
deviating up to $\sim$35\,s from the linear regression and allowing
for any possible value of the azimuthal order $m$.  
{ We plot echelle diagrams for the $\ell$=1 and $\ell$=2 sequences
in Fig.~6}, and list all $\ell$=1 and $\ell$=2 matches in Table
4.  Periods below 900 seconds that do match any of the $\ell$=1 or
$\ell$=2 sequences are indicated as such in Table 4 for completeness.

Obviously, the period spacing in \target\ is not very strict, which
makes conclusive identification of individual modes difficult, as
by-chance identifications of modes that do not belong to a sequence
are probable. However, the fact that the spacings vary within a
sequence may hold information about the internal structure that can be
derived from detailed seismic modelling.  From Fig.\ 6 we find that
between sequence numbers 8 and 18 in the $\ell$=1 sequence,
corresponding to pulsation periods between 3300--5900\,s, there is a
remarkable bump away from the average spacing.  { Whereas the $\ell$=1
sequence shows coherence over a large period range, the $\ell$=2
sequence is dominated by a clumping of multiplet candidates in the
period range 2900--3700\,s.}

Pulsation modes that could erroneously be accepted in any of the
sequences may stem from any of the following: modes with higher degree
$\ell$ that are strong enough to be picked-up by \kep; modes in
combination frequencies; modes that are tidally forced to have large
amplitudes; modes that originate in the companion.  If a mode in a
combination frequency is due to non-linear or resonance effects
between two other modes, its assignment to a period-spacing sequence
may not be justified. Therefore we mark the modes that take part in
frequency combinations (see Table 5) with a circle in Fig.\ 6.

\begin{figure}[h]
\centering
\psfig{figure=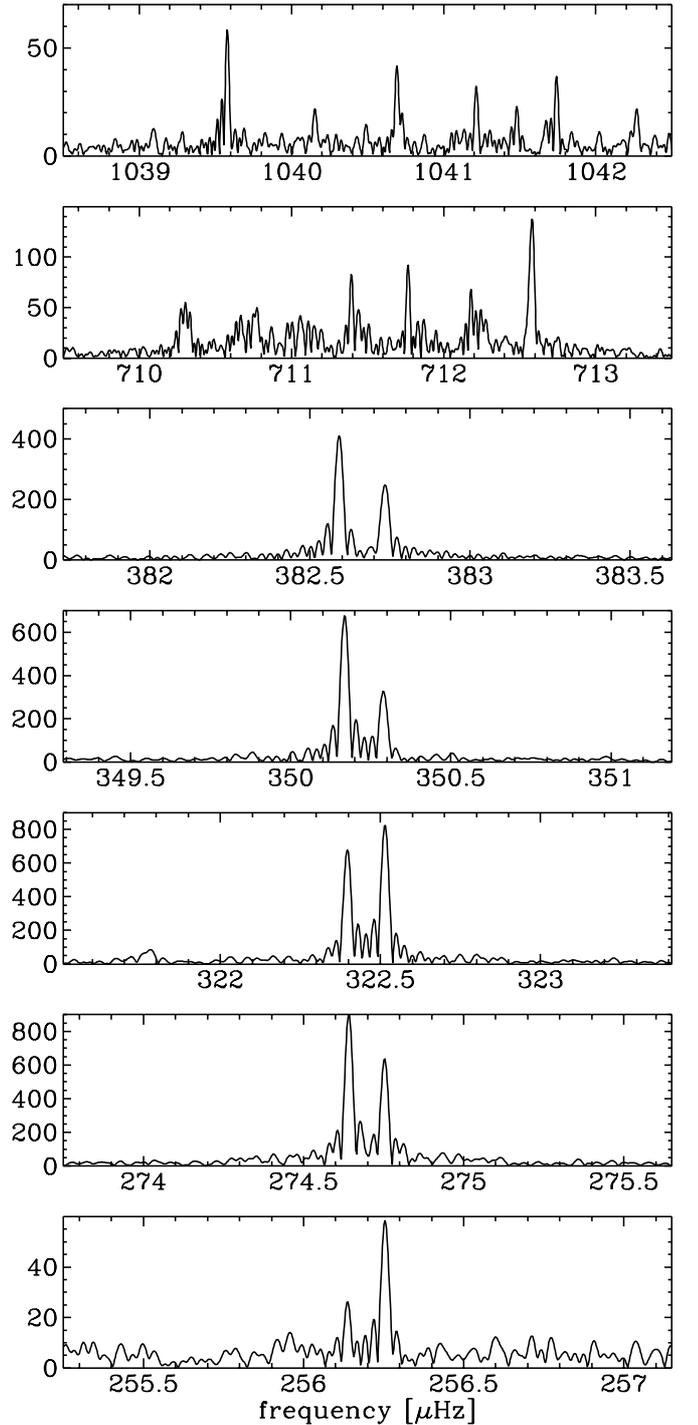,width=9cm}
\caption[]{Selected multiplets. The bottom 5 panels show the
  $\sim$0.13\,$\mu$Hz splitting seen in some of the $\ell$=1 doublets.
    The strongest pulsations we detect in \target\ are amongst the
    $\ell$=1 peaks shown here.
  The top panels show sample regions with sequences of frequencies
  with splittings in the 0.37--0.58\,$\mu$Hz range.  Amplitudes are
  given in ppm. The significance cut-off amplitude is 17\,ppm.}
\label{fig:multiplets}
\end{figure}

\begin{figure}[h]
\centering
\psfig{figure=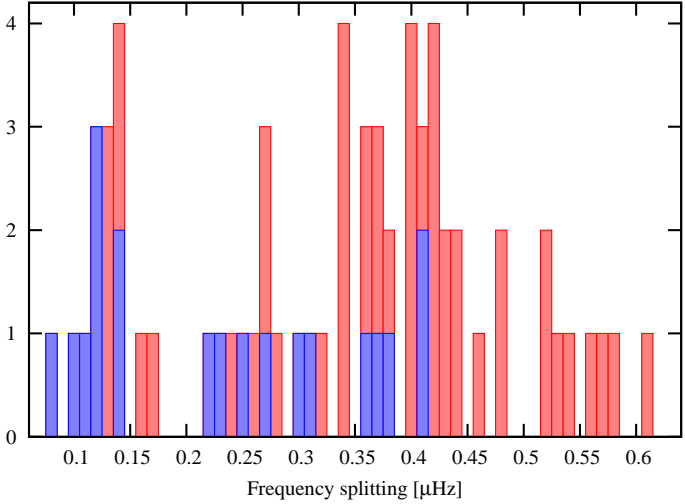,width=9cm}
\caption[]{Histogram of splittings listed in Table 4. The blue boxes
  reflect splittings from multiplets (mostly doublets) that were
  matched with the $\ell$=1 sequence in Sect.~6.}
\label{fig:splithisto}
\end{figure}

Tidally forced oscillations can have large amplitudes when close to an
eigenfrequency of the star, and can in principle be of any degree
$\ell$, although an $\ell$=2 excitation can be considered the most
probable, in which case tidally induced pulsations should be part of
the $\ell$=2 sequence. The error on the orbital frequency is not small
enough to exclude some of the strongest pulsations in Table 4 from
being a multiple of the orbital frequency.  We note however, that for
the observed frequencies in \target\ tidal forcing would occur at very
high harmonics of the orbital frequency { (i.e.~ $>$300 for the
  strongest pulsation frequencies)}, which may diminish the forcing
efficiency.  { Nevertheless, a recent case of tidally driven
  pulsations at high orbital harmonics (90th and 91st) is presented by
  \cite{welsh2011}.}

Even though pulsating white dwarfs are rare as the instability strips
are small, possible pulsations in the supposed white-dwarf companion
could be picked up by \kep.  Assuming a white dwarf with a large
radius { 0.013\,$R_{\sun}$} { ($\log(g)$=8.0, $M$=0.63\,$M_{\sun}$)},
with a similar or somewhat cooler temperature as the subdwarf, and
with intrinsic pulsation amplitudes at the 10--20\% level, we conclude
that such a white dwarf would contribute pulsation frequencies at
observable amplitudes to the frequency spectrum of \target, even if
the white dwarf is not detected in our spectra, or in the available
broad-band imaging. Hence we can
not exclude this possibility, when trying to match period
spacings or frequency splittings in \target.

If any of the significant peaks we have detected were to
originate from the white dwarf, their amplitudes would not be
maximised by shifting to the rest frame of the sdB, as we demonstrated
is the case for the short period peaks in Sect.~5.1. Rather, a R\o mer
delay with the opposite sign and an amplitude corresponding to the
radial velocity of the white dwarf would have to be imposed to recover
the full amplitude. Using the technique of Sect.~5.1 we found only one
peak in Table~4 that shows an orbital phase behaviour that is opposite
to that expected for the subdwarf, and hence could originate in the
supposed white-dwarf companion: it is the peak with the shortest
periodicity at 5053.5\,$\mu$Hz.  As this peak is just at the
detection limit we cannot rule out that it is just a random
effect. If it persists in the next year of \kep\ 
observations, it might turn out to be the first indication of an
sdB+WD system where both stars are pulsating.

\section {Frequency splittings and the rotation period}

Fixed frequency splittings are expected as a consequence of rotation.
To the first order in the rotation frequency $\Omega$, the observed
frequencies of the modes are altered by $m \Omega (1 - C_{n'l})$, with
$m$ the azimuthal quantum number of the spherical harmonic.  Hence,
for non-radial modes of given degree $\ell$, we expect frequency
multiplets of 2$\ell$+1 peaks to occur.  However, not all peaks of a
full multiplet need to be excited, and, furthermore, cancellation
effects for a given inclination angle can favour the observability of
some modes over others within a multiplet \citep[see e.g.][]{reed05}. 

In the case of \target\ the frequencies mostly do not appear as full
multiplets, which makes the interpreting of the observed splittings
difficult. See Table 4 for the observed splittings.  We find many
splittings in the range of 0.1--0.5\,$\mu$Hz.  In Fig.~7 we show two
sample frequency sequences that show a frequency splitting in the
0.4--0.55\,$\mu$Hz range, that seems typical for frequency groups that
do not match the $\ell$=1 period spacing.  We also show in Fig.\ 7 a
series of doublets that match the $\ell$=1 period spacing, which all
have a frequency splitting on the order of 0.13\,$\mu$Hz, and which in
fact cause the prominent beating seen in Fig.\ 4.  We also detect some
doublets with spacings approximately twice this ($\sim$0.23\,$\mu$Hz)
and we interpret the smaller frequency spacings as consecutive $m$
while the larger spacings would be the $|$$m$$|$=1 pairs.  { In
Fig.~8 we present a histogram of the splittings listed in Table 4.}

We tentatively identify this splitting of 0.13\,$\mu$Hz as the
rotation frequency of the subdwarf in \target.  We regard this result
as inconclusive as other doublets that seem to match the $\ell$=1
period spacing show a larger frequency splitting ($>$0.3\,$\mu$Hz). We
note however that some of the peaks in these $\ell$=1 doublets also
match the $\ell$=2 period spacing, and hence could be $\ell$=2 modes
instead.

Assuming that the Ledoux constant $C_{n'l}$ can be approximated by
1/($\ell$($\ell$+1)) as in the asymptotic $g$-mode regime, the
observed frequency splitting of 0.13\,$\mu$Hz in the $\ell$=1 doublets
leads to a rotation period of the subdwarf of 45\,d.  This
consequently implies that the rotation of the subdwarf is not
phase-locked with the orbit.

Even if we assume that the larger frequency splittings, in the
0.4--0.55\,$\mu$Hz range, belong to multiplets of $\ell$$>$1 modes and
are due to rotation, we would still conclude with certainty that the
subdwarf rotates subsynchronously with respect to the orbit.  Here the
case of a splitting of 0.55\,$\mu$Hz for a supposed $\ell$=2 multiplet
provides the lower limit to the rotation period of 18\,d.

\citet{pablo11,pablo12} discuss the frequency splittings in the three
of the four known \kep\ sdB+dM systems, which all three have orbital
periods on the order of ten hours.  From the frequency splittings they
derive that for each of their cases the subdwarf rotation is
subsynchronous with respect to the orbit.  Following their description
of the \citet{zahn75} synchronization timescale, we expect for \target\ a
much longer synchronization timescale than for those sdB+dM systems,
as the orbital period is much longer and the synchronization timescale
scales with $P_{\rm orb}^{17/3}$.  In fact, we expect that the
rotation period we measure now is a good indicator of the rotation
period of the subdwarf after it just settled on the extreme horizontal
branch.

\section{ Summary and Conclusions }

From our new low-resolution spectroscopy we discovered that
\target\ is a binary consisting of a B subdwarf and an unseen
companion, likely a white dwarf. We found a circular orbit with
$P_{\rm orb}$=10.05\,d, and a radial-velocity amplitude of
58km\,s$^{-1}$.  From the high signal-to-noise average spectrum we
redetermined the atmospheric parameters of the subdwarf:
\teff\,=\,27\,910\,K and \logg\,=\,5.41\,dex.

Five quarters of short-cadence \kep\ data of \target\ reveal Doppler
beaming at the 238 ppm level. The amplitude of the Doppler-beaming
modulation in the light curve corresponds to a radial-velocity
amplitude that is fully consistent with that determined from the
spectroscopic data.

We developed a new method that uses the pulsations as clocks to
measure the orbital light-travel effect, or R\o mer delay, directly
from the \kep\ light curve. The measured delay is again in
perfect agreement with the spectroscopically determined
radial-velocity amplitude.  This method allows the light-travel delay
to be measured on many pulsations simultaneously, allowing the
determination to be accurate, and hence can be useful for stars that
have many weak pulsation frequencies such as sdB stars.

For the 300 second $p$-modes in \target\ the pulsation amplitude
varies slightly when transforming the observation timings to that of
the center-of-mass of the binary.  This effect will only show up for
short-period pulsations in a long orbit.  We have been able to measure
the R\o mer delay from this effect.

One may wonder if it is still necessary to use spectroscopy to solve
orbits for targets in the \kep\ field.  For the case of \target\ the
phasing of the spectroscopic results with respect to the \kep\ light
curve proved essential in order to establish that the light-curve
modulation is due to Doppler beaming, and not due to reflected light
from the companion or to a contaminating object in the course
\kep\ pixel aperture.  { In general, the Doppler beaming amplitude
  may not reflect the radial-velocity amplitude in case the companion
  significantly contributes to the observed combined beaming
  amplitude. However, the phase and amplitude of the light-travel
  delay as determined by using the pulsations as clocks provide the
  same orbital constraints as derived from spectroscopy.}
 
We extracted 166 pulsation frequencies from the \kep\ light curve,
among which 13 are found to be possible combination frequencies.  We
found 6 $p$-modes, and 160 frequencies in the $g$-mode domain,
demonstrating the potential for a seismic analysis of this star.

We showed that many of the pulsation frequencies match period spacings
of $\ell$=1 and $\ell$=2 sequences, implying that the interior
structure of the subdwarf cannot be heavily stratified.  The
period-spacing sequences will aid in the identification process of
the modes in a future seismic study of this object.

Although we detect many pulsation frequencies, we see little 
evidence for complete multiplets.  The frequency splittings are many
in \target\ and are difficult to fit in a standard picture.  We
attribute the smallest splittings that we find to $\ell$=1 doublets,
and derive a rotation period of 45\,d.  As a lower limit for the
rotation period, we derive a value of 18\,d, which implies that the
subdwarf rotates subsynchronously with respect to the orbit.

\target\ is the first sdB pulsator in the \kep\ sample with a
confirmed compact stellar-mass companion. The 10 day orbit is in the
long end of the period range of the $\sim$100 known sdB binaries that
have periods compatible with the common-envelope ejection scenario.
Assuming a canonical sdB mass of 0.48\,$M_{\sun}$, we derived a lower
limit for the mass of the companion of 0.63\,$M_{\sun}$.  The distance
between the two companions is $\ga$20\,$R_{\sun}$, which implies that
if the sdB is a result of a common-envelope phase the progenitor must
have been close to the maximum radius for a red giant, near the tip of
the red-giant branch.

\acknowledgements

Based on observations made with the Nordic Optical Telescope, operated
on the island of La Palma jointly by Denmark, Finland, Iceland,
Norway, and Sweden, in the Spanish Observatorio del Roque de los
Muchachos (ORM) of the Instituto de Astrofisica de Canarias, and the
William Herschel Telescope and Isaac Newton Telescope also at ORM,
operated by the Isaac Newton Group.   

MDR and LF were Visiting Astronomers to the Kitt Peak National
Observatory, National Optical Astronomy Observatory, which is operated
by the Association of Universities for Research in Astronomy (AURA)
under cooperative agreement with the National Science Foundation.

MDR and LF were supported by the Missouri Space Grant Consortium
funded by NASA.

The authors gratefully acknowledge the \kep\ team and all who
have contributed to enabling the mission. Funding for the \kep\ 
Mission is provided by NASA's Science Mission Directorate.

The research leading to these results has received funding from the
European Research Council under the European Community's Seventh
Framework Programme (FP7/2007--2013)/ERC grant agreement
N$^{\underline{\mathrm o}}$\,227224 ({\sc prosperity}), as well as
from the Research Council of KU~Leuven grant agreement GOA/2008/04.

JHT cordially thanks the Instituut voor
Sterrenkunde, KU~Leuven, for its hospitality.
\bibliographystyle{aa}
\bibliography{sdbrefs}

\end{document}